\documentclass[12pt,fleqn,a4paper]{article}
\usepackage{amssymb}
\usepackage{amsmath}
\usepackage[dvips]{graphicx,psfrag}
\newcommand{\psla}{p\hspace{-0.55em}/}
\newcommand{\qsla}{q\hspace{-0.55em}/}

\newcommand{\ssla}{s\hspace{-0.55em}/}
\newcommand{\epssla}{\epsilon\hspace{-0.45em}/}
\begin{document}
\begin{flushright}
KEK preprint 2000-131 \\
KEK-TH-732 \\
{\tt hep-ph/0012040 }\\
December 2000 \\
\end{flushright}
\vspace*{1cm}
\begin{center}
{\baselineskip 25pt
\large{\bf 
P and T Odd Asymmetries
in Lepton Flavor Violating $\tau$ Decays
}}

\vspace{1cm}

{\large Ryuichiro Kitano\footnote
{email: {\tt ryuichiro.kitano@kek.jp}}
and Yasuhiro Okada\footnote
{email: {\tt yasuhiro.okada@kek.jp}}
}
\vspace{.5cm}

{\small {\it Theory Group, KEK, Oho 1-1, Tsukuba, 
Ibaraki 305-0801, Japan \\
and \\
Department of Particle and Nuclear Physics,
The Graduate University for Advanced Studies,\\
Oho 1-1, Tsukuba, Ibaraki 305-0801, Japan}}

\vspace{1.5cm}
{\bf Abstract}
\end{center}

\bigskip
We calculated the 
differential cross sections
of the processes in which
one of the pair created $\tau$ particles
at an $e^+  e^-$ collider 
decays into lepton flavor violating final states
e.g.\ $\tau \to \mu \gamma$, $\tau \to 3 \mu$, $\tau \to \mu ee$.
Using the correlations between angular
distributions of both sides of $\tau$ decays,
we can obtain information on
parity and CP violations of lepton flavor 
non-conserving interactions.
The formulae derived here are
useful in distinguishing different 
models,
since each model of physics
beyond the standard model predicts
different angular correlations.
We also calculate angular distributions
of the major background process 
to $\tau \to l \gamma$ search,
namely $\tau \to l \nu \bar{\nu} \gamma$,
and discuss usefulness of 
the angular correlation
for background suppression.

\newpage

\section{Introduction}
\baselineskip 20pt

Recent results form neutrino experiments
such as the Super-Kamiokande experiment
strongly suggest neutrino oscillation 
so that
there are flavor mixings
in the lepton sector \cite{Fukuda:1998mi}.
This implies that
the charged lepton flavor violating processes
such as $\mu \to e \gamma$, $\tau \to \mu \gamma$,
$\tau \to e \gamma$, etc.\ also
occur at some level.
It is, therefore,
important to search for lepton flavor violation (LFV)
in rare decay processes of muon and $\tau$.

Prediction of branching fraction of
LFV processes depends on
models of physics beyond the standard model.
In the minimal extension of the standard model
which takes into account neutrino oscillations
by seesaw mechanism 
of neutrino mass generation \cite{seesaw},
the expected branching fraction is 
too small to be observable in near future \cite{Cheng:1980tp}.
On the other hand,
in supersymmetric (SUSY) models,
the prediction can be close to the 
current experimental upper bound.
In this case,
the flavor mixing in the
slepton mass matrix becomes a new source of LFV.
Even in the minimal supergravity scenario \cite{Chamseddine:1982jx},
in which the slepton mass matrix is proportional to
a unit matrix at the Planck scale,
the renormalization effects
due to LFV interactions
can induce sizable slepton mixings \cite{Hall:1986dx}.
For example, such LFV Yukawa interactions exist
in SUSY Grand Unified Theory (GUT) \cite{Barbieri:1994pv},
SUSY model with right-handed neutrinos \cite{Borzumati:1986qx},
and SUSY models with
exotic vectorlike leptons \cite{Kitano:2000zw}.
Another interesting possibility is 
models with extra dimensions, 
where
the neutrino masses and mixings are
obtained from the Yukawa interaction
between the ordinary left-handed leptons and
the gauge-singlet neutrinos which propagate 
in the bulk of extra dimensions \cite{Arkani-Hamed:1998vp}.
This Yukawa interaction breaks
the lepton flavor conservation and
the Kaluza-Klein modes of the bulk neutrinos
can enhance $\mu \to e \gamma$ decay, $\tau \to \mu \gamma$ decay, etc.\
through the loop diagrams \cite{Faraggi:1999bm}.

In the muon decay,
the polarized muon experiments
provide useful information on the 
nature of LFV interactions \cite{Okada:1998fz, Okada:2000zk}.
We can define a parity (P) odd asymmetry
for $\mu \to e \gamma$ process and 
P and time reversal (T) odd asymmetries for $\mu \to 3e$ process.
These asymmetries are useful to 
distinguish different models.
For example,
in the SU(5) SUSY GUT model
with small and intermediate value of $\tan \beta$
(a ratio of two vacuum expectation values of Higgs fields),
only $\mu^+ \to e^+_L \gamma$ (or $\mu^- \to e^-_R \gamma$)
occurs because LFV is induced through
the right-handed slepton sector.
On the other hand,
the SUSY models with right-handed neutrinos
predict $\mu^+ \to e^+_R \gamma$ (or $\mu^- \to e^-_L \gamma$),
and the SUSY models with vectorlike leptons
can induce both $\mu^+ \to e^+_L \gamma$ 
and $\mu^+ \to e^+_R \gamma$
depending on how interaction breaks the
lepton flavor conservation.
In the models with extra dimensions,
only $\mu^+ \to e^+_R \gamma$ can occur.
As for the T odd asymmetry in the $\mu \to 3e$ process,
it was shown that asymmetry could be sizable
in the SU(5) SUSY GUT \cite{Okada:1998fz, Okada:2000zk}.

In this paper,
we discuss the LFV processes of
$\tau$ decays such as $\tau \to \mu \gamma$,
$\tau \to 3 \mu$,
$\tau \to \mu ee$,
taking into account P and T odd asymmetries.
In the $\tau^+ \tau^-$ pair production
at $e^+ e^-$ collision,
we can extract information on the spin
of the decaying $\tau$ particle
from the angular distribution of the 
$\tau$ decay products
in the opposite side.
Using this technique, 
we can obtain the P and T odd asymmetry
defined in the rest frame of $\tau$.
The method of the spin correlation has
been developed since the days before
the discovery of $\tau$ particle \cite{Tsai:1971vv}.
There have been many works on spin correlation method
in search for anomalous coupling involving 
$\tau$ \cite{Kuhn:1984di}.
We have applied the formalism in
order to obtain the information on LFV interactions
under P and T symmetries.
We also calculate angular correlation
of the process where one of the $\tau$'s decays
through $\tau \to l \nu \bar{\nu} \gamma$ mode.
This mode is a background process 
to the $\tau \to l \gamma$ search
if the neutrinos carry out little energy.
As in the muon case \cite{Kuno:1996kv},
the angular correlation is 
useful to identify the
background process and the 
background suppression is 
effective for $\tau^- \to \mu^-_L \gamma$
($\tau^+ \to \mu^+_R \gamma$) search.

This paper is organized as follows.
In section 2,
we introduce a formalism to calculate the
spin correlation.
In section 3,
we present a differential cross section
of the production and decays of $\tau^+ \tau^-$
at $e^+ e^-$ colliders
where one of $\tau$'s decays in $\tau \to \mu \gamma$
or $\tau \to e \gamma$ modes,
and show how to extract P-odd asymmetry of $\tau$ decay.
In section 4,
P and T odd asymmetries in three body LFV decays 
($\tau \to 3 \mu$, $\tau \to \mu ee$, etc.\ )
are considered.
In section 5,
we consider $\tau \to l \nu \bar{\nu} \gamma$ mode and
show that the analysis of the angular distribution
is useful for background suppression of the 
$\tau \to \mu \gamma$ and $\tau \to e \gamma$ searches.
A summary is given in section 6.
Appendices contain
the derivation of basic formulae and 
a list of kinematical functions.

\section{General formula for spin correlation}

In this section,
we present general formulae used 
in the calculation of 
differential cross sections and spin correlations.

We calculate differential cross sections of 
$e^+ e^- \to \tau^+ \tau^- \to f_B f_A$,
where $f_B$ ($f_A$) represents the decay products of 
$\tau^+$ ($\tau^-$).
If the intermediate states were spinless particles,
the cross section is simply a product of
a production cross section and decay branching ratios.
However,
in the case of spin $1/2$ particles,
we have to take into account spin correlation
between two intermediate particles.
If we take $\tau^+ \to f_B$ to be a LFV decay mode,
we can measure P and T violation
of LFV interactions 
by using the angular correlations of 
decay products of $\tau^+$ and $\tau^-$.

The differential cross section of 
$e^+ e^- \to \tau^+ \tau^- \to f_B f_A$ is given by
\begin{eqnarray}
 d \sigma = d\sigma^P \ dB^{\tau^- \to f_A}\ dB^{\tau^+ \to f_B} +
\sum_{a,b=1}^3
d\Sigma^P_{ab} \ dR^{\tau^- \to f_A}_a \ 
dR^{\tau^+ \to f_B}_b\ ,
\label{neo1}
\end{eqnarray}
and 
\begin{eqnarray}
 d \sigma^P = \frac{d^3 p_A}{(2 \pi)^3 2 p_A^0}\ 
\frac{d^3 p_B}{(2 \pi)^3 2 p_B^0}\ 
\frac{1}{2s}\ 
( 2 \pi )^4 \ \delta^4 (p_A + p_B -p_{e^+} - p_{e^-})\ 
\alpha^P\ ,
\end{eqnarray}
\begin{eqnarray}
 dB^{\tau^- \to f_A} =
\frac{1}{\Gamma}\ 
\frac{d^3 q_1}{(2 \pi)^3 2 q_1^0} \cdots 
\frac{d^3 q_n}{(2 \pi)^3 2 q_n^0} \ 
\frac{1}{2 m_\tau}\ 
( 2 \pi )^4 \ \delta^4 \left( \sum_{i=1}^n q_i - p_A \right)\ 
\alpha^{D_-}\ ,
\end{eqnarray}
\begin{eqnarray}
 dB^{\tau^+ \to f_B} =
\frac{1}{\Gamma}\ 
\frac{d^3 q_{n+1}}{(2 \pi)^3 2 q_{n+1}^0} \cdots
\frac{d^3 q_{n+m}}{(2 \pi)^3 2 q_{n+m}^0} \  
\frac{1}{2 m_\tau} \ 
( 2 \pi )^4 \  \delta^4 \left( \sum_{i=n+1}^{n+m} q_i - p_B \right)\ 
\alpha^{D_+}\ ,
\end{eqnarray}
\begin{eqnarray}
 d \Sigma_{ab}^P = \frac{d^3 p_A}{(2 \pi)^3 2 p_A^0}\ 
\frac{d^3 p_B}{(2 \pi)^3 2 p_B^0}\ 
\frac{1}{2s}\ 
( 2 \pi )^4 \ \delta^4 (p_A + p_B -p_{e^+} - p_{e^-})\ 
\rho^P\ ,
\end{eqnarray}
\begin{eqnarray}
 dR_a^{\tau^- \to f_A} =
\frac{1}{\Gamma}\ 
\frac{d^3 q_1}{(2 \pi)^3 2 q_1^0} \cdots 
\frac{d^3 q_n}{(2 \pi)^3 2 q_n^0} \ 
\frac{1}{2 m_\tau}\ 
( 2 \pi )^4 \ \delta^4 \left( \sum_{i=1}^n q_i - p_A \right)\ 
\rho_a^{D_-}\ ,
\end{eqnarray}
\begin{eqnarray}
 dR_b^{\tau^+ \to f_B} =
\frac{1}{\Gamma}\ 
\frac{d^3 q_{n+1}}{(2 \pi)^3 2 q_{n+1}^0} \cdots
\frac{d^3 q_{n+m}}{(2 \pi)^3 2 q_{n+m}^0} \  
\frac{1}{2 m_\tau} \ 
( 2 \pi )^4 \  \delta^4 \left( \sum_{i=n+1}^{n+m} q_i - p_B \right)\ 
\rho_b^{D_+}\ ,
\end{eqnarray}
where we assume that
$f_A$ is a $n$ body system and 
$f_B$ is a $m$ body system.
$p_{e^+}$ ($p_{e^-}$) is $e^+$ ($e^-$) four momentum,
$p_B$ ($p_A$) is $\tau^+$ ($\tau^-$) four momentum,
and $q_i$'s are momenta of 
final state particles.
$s$ is determined as
$s= (p_{e^+} + p_{e^-})^2$.
$\Gamma$ and $m_\tau$ are the width 
and the mass of the $\tau$, respectively.
In order to define $\alpha^P$, $\alpha^{D_-}$,
$\alpha^{D_+}$, $\rho^P_{ab}$, $\rho^{D_-}_a$,
and $\rho^{D_+}_b$,
we first write down the invariant
amplitude of $e^+ e^- \to \tau^+ \tau^- \to f_B f_A$ 
as follows:
\begin{eqnarray}
 M \!\!\!&=&\!\!\!
\frac{e^2}{s}\ 
\bar{A} ( \psla_A + m_\tau ) \gamma^\mu ( \psla_B - m_\tau ) B
\
\bar{v}_{e^+} \gamma_\mu u_{e^-}
\
\nonumber \\
&&\!\!\!
\times 
\frac{1}{p_A^2 - \left( m_\tau - \frac{i \Gamma}{2} \right)^2 }
\ 
\frac{1}{p_B^2 - \left( m_\tau - \frac{i \Gamma}{2} \right)^2 }
\ ,
\label{1}
\end{eqnarray}
where 
$v_{e^+}$ ($u_{e^-}$) is the wave function of 
positron (electron) and 
$A$ and $B$ are spinors which include
wave functions of final states and interaction vertices.
By using the Bouchiat-Michel formulae \cite{bouchiat:1958}
and
the narrow width approximation,
$\alpha^P$, $\alpha^{D_-}$,
$\alpha^{D_+}$, $\rho^P_{ab}$, $\rho^{D_-}_a$,
and $\rho^{D_+}_b$ are
given by
\begin{eqnarray}
 \alpha^P =
\frac{1}{4} \  \frac{e^4}{s^2} \ 
{\rm Tr} \left[
( \psla_A + m_\tau ) \gamma^\mu
( \psla_B - m_\tau ) \gamma^\nu
\right]\ 
{\rm Tr} \left[
\psla_{e^+} \gamma_\mu \psla_{e^-} \gamma_\nu
\right]\ ,
\label{3}
\end{eqnarray}
\begin{eqnarray}
 \alpha^{D_-} =
\frac{1}{2}\ 
\left \{
\bar{A} ( \psla_A + m_\tau ) A
\right \}\ ,
\label{4}
\end{eqnarray}
\begin{eqnarray}
 \alpha^{D_+} =
\frac{1}{2}\ 
\left \{
\bar{B} ( \psla_B - m_\tau ) B
\right \}\ ,
\label{5}
\end{eqnarray}
\begin{eqnarray}
 \rho^P_{ab} =
\frac{1}{4} \  \frac{e^4}{s^2} \ 
{\rm Tr} \left[
\gamma_5 \  \ssla_A^a \ ( \psla_A + m_\tau ) \gamma^\mu
\gamma_5 \  \ssla_B^b \ ( \psla_B - m_\tau ) \gamma^\nu
\right]\ 
{\rm Tr} \left[
\psla_{e^+} \gamma_\mu \psla_{e^-} \gamma_\nu
\right]\ ,
\label{6}
\end{eqnarray}
\begin{eqnarray}
 \rho^{D_-}_a =
\frac{1}{2}\ 
\left \{
\bar{A} \ \gamma_5\  \ssla_A^a \ ( \psla_A + m_\tau ) A
\right \}\ ,
\label{7}
\end{eqnarray}
\begin{eqnarray}
 \rho^{D_+}_b =
\frac{1}{2}\ 
\left \{
\bar{B} \ \gamma_5 \  \ssla_B^b \ ( \psla_B - m_\tau ) B
\right \}\ ,
\label{8}
\end{eqnarray}
where the spins of the final state fermions
are summed over,
and four vectors $(s_A^a)^\mu$ and $(s_B^b)^\nu$ 
($a,b=1,2,3$)
are 
a set of vectors which satisfy following equations.
\begin{eqnarray}
&&  p_A \cdot s_A^a = p_B \cdot s_B^b = 0, \label{9}\\
&&  s_A^a \cdot s_A^b = s_B^a \cdot s_B^b = - \delta^{ab} \ , \label{10}
\\
&&  \sum_{a=1}^3 (s_A^a)_\mu (s_A^a)_\nu = 
- g_{\mu \nu} 
+ \frac{p_{A \mu} p_{A \nu}}{m_\tau^2}\ ,\ \ \ 
\sum_{b=1}^3 (s_B^b)_\mu (s_B^b)_\nu = 
- g_{\mu \nu} 
+ \frac{p_{B \mu} p_{B \nu}}{m_\tau^2}\ .
\label{11}
\end{eqnarray}
The derivation of the above result
is shown in Appendix \ref{formula}.
Notice that
$d\sigma^P$, $dB^{\tau^- \to f_A}$, and $dB^{\tau^+ \to f_B}$
in eq.(\ref{neo1})
are the $\tau^+ \tau^-$ production
cross section and $\tau$ decay branching ratios,
in which spins of $\tau$'s are averaged,
and 
$d\Sigma_{ab}^P$, $dR_a^{\tau^- \to f_A}$,
and $dR_b^{\tau^+ \to f_B}$
represent the spin correlation effects of this process.

In above formulae,
it is assumed that the $\tau$ pair production
occurs through the photon exchange.
It is straightforward to include the 
contribution from
the $Z$ boson exchange and the $\gamma-Z$ interference.
If we consider the $e^+ e^-$ center of mass energy
to be in the range of the $\tau^+ \tau^-$ threshold 
energy considered in the $\tau$-charm factory or the
$\Upsilon (4S)$ resonance energy where
$e^+ e^-$ B-factories are operated,
these effects only contribute to the 
production cross section at the level of
$O(10^{-4})$ of the photon exchanging diagram.

\section{Parity asymmetry in $\tau \to \mu \gamma$ decay}

Let us calculate the 
cross section of 
$e^+ e^- \to \tau^+ \tau^- \to \mu^+ \gamma + f_A$ processes.
For $f_A$,
we consider hadronic and leptonic modes such as
($\pi \nu$, $\rho \nu$, $a_1 \nu$, and $l \bar{\nu} \nu$).
Below we neglect the muon mass compared to $\tau$ mass,
and therefore all formulae can be applied also to the 
$\tau \to e \gamma$ process.
The effective Lagrangian for $\tau^+ \to \mu^+ \gamma$ decay
is given by
\begin{eqnarray}
 {\cal L}=
-\frac{4 G_{\rm F}}{\sqrt{2}}
\left \{
m_\tau A_R \bar{\tau} \sigma^{\mu \nu} P_L \mu F_{\mu \nu} +
m_\tau A_L \bar{\tau} \sigma^{\mu \nu} P_R \mu F_{\mu \nu} + {\rm h.c.}
\right \}\ ,
\label{13}
\end{eqnarray}
where $G_{\rm F}$ is the Fermi coupling constant,
$P_L = (1-\gamma_5)/2$, and $P_R = (1+\gamma_5)/2$.
The operator with the coupling constant $A_R$ ($A_L$)
induces the $\tau^+ \to \mu^+_R \gamma$ 
($\tau^+ \to \mu^+_L \gamma$) decay.
As mentioned in Introduction,
each model 
of the physics beyond the standard model
predicts a different ratio of $A_L$ and $A_R$.
For example,
the SU(5) SUSY GUT
in the minimal supergravity scenario
predicts that only $A_L$ has a non-vanishing value
for small and intermediate values of $\tan \beta$.
Therefore the separate determination of $A_L$ and $A_R$
provides us important information on
the origin of LFV.
For this purpose,
we need information about the $\tau$ polarization.
This can be done by observing angular distributions
of final state of $\tau$ decay in the opposite side
in the modes of 
$\tau \to \pi \nu$, $\tau \to \rho \nu$, $\tau \to a_1 \nu$,
and $\tau \to l \bar{\nu} \nu$,
because these processes proceed due to the $V-A$ interaction
and therefore have a specific angular distribution
with respect to polarization of $\tau$.
Using $\tau^+ - \tau^-$ spin correlation,
we can determine $|A_L|^2$ and $|A_R|^2$, separately.

\begin{figure}
\psfrag{e^+}{$e^+$}
\psfrag{e^-}{$e^-$}
\psfrag{tau^+}{$\tau^+$}
\psfrag{tau^-}{$\tau^-$}
\psfrag{th_tau}{$\theta_\tau$}
\psfrag{x_1}{$x_1$}
\psfrag{y_1}{$y_1$}
\psfrag{z_1}{$z_1$}
\psfrag{x_2}{$x_2$}
\psfrag{y_2}{$y_2$}
\psfrag{z_2}{$z_2$}
\psfrag{x_3}{$x_3$}
\psfrag{y_3}{$y_3$}
\psfrag{z_3}{$z_3$}
\psfrag{Frame 1}{Frame 1}
\psfrag{Frame 2}{Frame 2}
\psfrag{Frame 3}{Frame 3}
\begin{center}
\includegraphics[width=10cm]{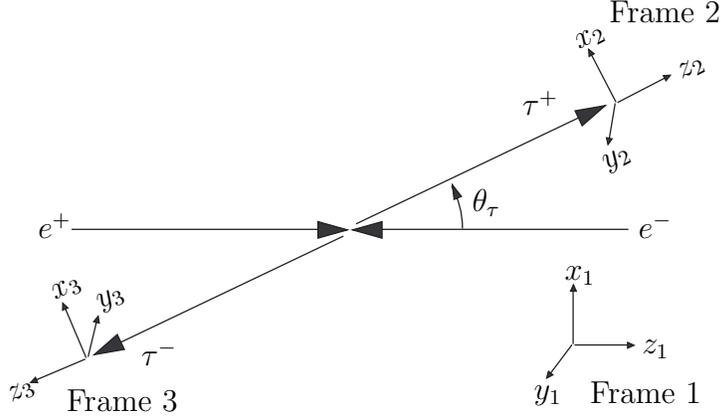} 
\end{center}
\caption{The coordinate systems.
The plane determined by $e^+ e^-$ and $\tau^+ \tau^-$ 
momentum vectors corresponds to the $xz$-planes
in each of three coordinate systems.}
\label{fig1}
\end{figure}

We first define three coordinate systems (Fig.\ref{fig1}).
The first coordinate system (Frame 1)
is the center of mass frame of the $e^+ e^-$ collision
in which the $z$ axis is taken to be the $e^+$ momentum direction.
The second one (Frame 2) is the rest frame 
of the $\tau^+$,
and the
third one (Frame 3) is
the rest frame of the $\tau^-$.
More explicitly,
the relation of a four vector
in the three systems are given as follows:
\begin{eqnarray}
\xi^\mu_1 \!\!\!&=&\!\!\!
\left(
\begin{array}{cccc}
 1 & 0 & 0 & 0 \\
 0 & \cos \theta_\tau & 0 & \sin \theta_\tau  \\
 0 & 0 & 1 & 0 \\
 0 & -\sin \theta_\tau & 0 & \cos \theta_\tau \\
\end{array}
\right)
\left(
\begin{array}{cccc}
 \gamma & 0 & 0 & \gamma \beta_\tau \\
 0 & 1 & 0 & 0 \\
 0 & 0 & 1 & 0 \\
 \gamma \beta_\tau & 0 & 0 & \gamma \\
\end{array}
\right)
\xi^\mu_2 \nonumber \\
\!\!\!&=&\!\!\!
\left(
\begin{array}{cccc}
 1 & 0 & 0 & 0 \\
 0 & 1 & 0 & 0 \\
 0 & 0 & -1& 0 \\
 0 & 0 & 0 & -1 \\
\end{array}
\right)
\left(
\begin{array}{cccc}
 1 & 0 & 0 & 0 \\
 0 & \cos \theta_\tau & 0 & -\sin \theta_\tau \\
 0 & 0 & 1 & 0 \\
 0 & \sin \theta_\tau & 0 & \cos \theta_\tau \\
\end{array}
\right)
\left(
\begin{array}{cccc}
 \gamma & 0 & 0 & \gamma \beta_\tau \\
 0 & 1 & 0 & 0 \\
 0 & 0 & 1 & 0 \\
 \gamma \beta_\tau & 0 & 0 & \gamma \\
\end{array}
\right)
\xi_3^\mu
\ ,
\label{14}
\end{eqnarray}
where $\gamma= \sqrt{s}/(2 m_\tau)$ 
and $\beta_\tau = \sqrt{ 1- 4m_\tau^2/s}$,
and the four vectors $\xi_{1-3}$ are 
defined in Frame 1-3, respectively.
We calculate
the production process in Frame 1,
and
the $\tau^+$ ($\tau^-$) decay in Frame 2 (Frame 3).
In the calculations,
we choose the spin vectors $(s_A^a)^\mu$, $(s_B^a)^\mu$
as follows:
\begin{eqnarray}
 (s_A^a)^\mu = \left(
\begin{array}{c}
 0 \\
 \delta_{a \mu} \\
\end{array}
\right) \ \ \ {\rm (in\ Frame\ 3)}\ ,
\label{neo15}
\end{eqnarray}
\begin{eqnarray}
  (s_B^b)^\mu = \left(
\begin{array}{c}
 0 \\
 \delta_{b \mu} \\
\end{array}
\right) \ \ \ {\rm (in\ Frame\ 2)}\ .
\label{16}
\end{eqnarray}

The production cross section and 
spin dependence term are
obtained from eq.(\ref{3}) and eq.(\ref{6}) as follows:
\begin{eqnarray}
 d\sigma^P \!\!\!&=&\!\!\!
\frac{d \Omega_{\tau}}{4 \pi}\ 
\frac{\pi \alpha^2}{s} \sqrt{1-\frac{4 m_\tau^2}{s}}
\left \{ 
\left(
1+\frac{4 m_\tau^2}{s}
\right) +
\left(
1- \frac{4 m_\tau^2}{s}
\right) \cos^2 \theta_\tau
\right \}\ ,
\\
d\Sigma^P_{ab} \!\!\!&=&\!\!\!
\frac{d \Omega_{\tau}}{4 \pi}\ 
\frac{\pi \alpha^2}{s} \sqrt{1-\frac{4 m_\tau^2}{s}}
\nonumber \\
&& \!\!\! \times
\left(
\begin{array}{ccc}
 \left( 1+ \frac{4 m_\tau^2}{s} \right) \sin^2 \theta_\tau & 0 
 & -\frac{2 m_\tau}{\sqrt{s}} \sin 2 \theta_\tau \\
 0 & \left( 1- \frac{4 m_\tau^2}{s} \right) \sin^2 \theta_\tau & 0 \\
 \frac{2 m_\tau}{\sqrt{s}} \sin 2 \theta_\tau & 0 
 & - \left( 1- \frac{4 m_\tau^2}{s} \right) - 
\left( 1+ \frac{4 m_\tau^2}{s} \right)
\cos^2 \theta_\tau \\
\end{array}
\right)\ .
\nonumber \\
\end{eqnarray}
where $\theta_\tau$ is the angle between the $e^+$ and $\tau^+$
directions in the Frame 1,
and $d \Omega_\tau$ is a solid angle element of $\tau^+$,
$d \Omega_\tau = d \cos \theta_\tau \ d \phi_\tau$.

For decay processes,
we take $\tau^+ \to \mu^+ \gamma$ for the $\tau^+$
side and
hadronic 
($\tau^- \to \pi^- \nu$, $\tau^- \to \rho^- \nu$, 
and $\tau^- \to a_1 \nu$)
and leptonic
($\tau^- \to l^- \bar{\nu} \nu$)
decays for the $\tau^-$ side.
$dB^{\tau^+ \to \mu^+ \gamma}$
and $dR^{\tau^+ \to \mu^+ \gamma}_{b}$
(see eq.(\ref{neo1}))
can be calculated from
eq.(\ref{5}) and eq.(\ref{8}) 
in which the spinor $B$ is given by 
\begin{eqnarray}
B=\frac{8i}{\sqrt{2}}\ G_{\rm F} m_\tau \sigma^{\mu \nu} 
(q_\gamma)_\mu (A_R P_L + A_L P_R ) \epsilon_\nu^* v(q_\mu)
\ ,
\label{15}
\end{eqnarray}
where $\epsilon_\nu$ is the polarization vector
of the photon 
and $(q_\gamma)_\mu$ is the momentum of the photon
and $v(q_\mu)$ is the wave function of the muon.
These quantities are given as follows:
\begin{eqnarray}
 dB^{\tau^+ \to \mu^+ \gamma} =
\frac{d \Omega_\mu}{4 \pi}\ 
\frac{1}{\Gamma}\ \frac{2}{\pi}\ 
G_{\rm F}^2 m_\tau^5\ 
(|A_L|^2 + |A_R|^2)
\ ,
\label{18}
\end{eqnarray}
\begin{eqnarray}
 dR^{\tau^+ \to \mu^+ \gamma}_b =
 \frac{d \Omega_\mu}{4 \pi}\ 
\frac{1}{\Gamma}\ \frac{2}{\pi}\ 
G_{\rm F}^2 m_\tau^5\ 
(|A_L|^2 - |A_R|^2)\ 
\left(
\begin{array}{c}
 \sin \theta_\mu \cos \phi_\mu \\
 \sin \theta_\mu \sin \phi_\mu \\
 \cos \theta_\mu \\
\end{array}
\right)\ ,
\label{19}
\end{eqnarray}
where $(\theta_\mu, \phi_\mu)$ 
are angles in the polar coordinate for a unit vector
of the muon momentum direction in Frame 2.
The three components in eq.(\ref{19}) corresponds to
$b=1,2,$ and $3$.

Next, we list $dB$ and $dR$ for the 
$\tau^-$ decay in each mode of $\tau^- \to \pi^- \nu$,
$\tau^- \to \rho^- \nu$, $\tau^- \to a_1^- \nu$,
and $\tau^- \to l^- \bar{\nu} \nu$.
For $\tau^- \to \pi^- \nu$ decay,
the spinor $A$ in eqs.({\ref{4}}) and (\ref{7}) is given by
\begin{eqnarray}
 A = 2 i V_{ud} f_\pi G_{\rm F} \qsla_\pi P_L u(q_\nu)\ ,
\end{eqnarray}
where $f_\pi$ is the pion decay constant,
$q_\pi$ is the momentum of the pion, and
$u(q_\nu)$ is the neutrino wave function.
Then, $dB^{\tau^- \to \pi^- \nu}$
and $dR^{\tau^- \to \pi^- \nu}_a$
are given by
\begin{eqnarray}
 dB^{\tau^- \to \pi^- \nu} =
\frac{d\Omega_{\pi}}{4 \pi}\ 
\frac{1}{\Gamma}\ 
\frac{1}{8 \pi}\ 
|V_{ud}|^2 f_\pi^2 G_{\rm F}^2 m_\tau^3\ ,
\label{28}
\end{eqnarray}
\begin{eqnarray}
 dR^{\tau^- \to \pi^- \nu}_a =
 dB^{\tau^- \to \pi^- \nu} 
\left(
\begin{array}{c}
 \sin \theta_\pi \cos \phi_\pi \\
 \sin \theta_\pi \sin \phi_\pi \\
 \cos \theta_\pi \\
\end{array}
\right)
\ ,
\label{26}
\end{eqnarray}
where ($\theta_\pi, \phi_\pi$) are the polar angles of
$\pi^-$ momentum in Frame 3 
and
$d \Omega_\pi = d \cos \theta_\pi \ d\phi_\pi$
(we use a similar notation in the following expressions).
Here
we neglect the mass of 
the pion. 
As before three elements in eq.(\ref{26}) corresponds to
$a=1,2,$ and $3$.
Similar results can be obtained for 
the vector mesons.
The spinor $A$ for $\tau \to \rho \nu$, $\tau \to a_1 \nu$
is given by 
\begin{eqnarray}
 A = - 2 V_{ud} g_V G_{\rm F} \epssla_V P_L u(q_\nu) \ .
\end{eqnarray}
where $g_V$ and $\epsilon_V$ are the decay constant 
and polarization vector 
of the corresponding vector mesons, respectively.
From this expression,
we can obtain $dB$ and $dR$ for 
the longitudinally polarized vector mesons
e.g.\ $\tau^- \to \rho^- (L) \nu$ and 
$\tau^- \to a_1^- (L) \nu$ as follows:
\begin{eqnarray}
 dB^{\tau^- \to V(L)^- \nu} \!\!\!&=&\!\!\!
\frac{d\Omega_{V}}{4 \pi}\ 
\frac{1}{\Gamma}\ 
\frac{1}{8 \pi}\ 
|V_{ud}|^2 \left( \frac{g_V}{m_V^2} \right)^2
G_{\rm F}^2 m_\tau^4 m_V^2 (1-\frac{m_V^2}{m_\tau^2})\ ,
\\
dR^{\tau^- \to V(L)^- \nu}_a \!\!\!&=&\!\!\!
dB^{\tau^- \to V(L)^- \nu}
\left(
\begin{array}{c}
 \sin \theta_V \cos \phi_V \\
 \sin \theta_V \sin \phi_V \\
 \cos \theta_V \\
\end{array}
\right)\ ,
\end{eqnarray}
where $m_V$ and ($\theta_V, \phi_V$) are 
the mass, and polar angles of the corresponding vector meson,
respectively.
For the transversely polarized vector mesons,
the spin dependence terms have a minus sign
contrary to the case of
the pion and longitudinally polarized vector mesons.
\begin{eqnarray}
 dB^{\tau^- \to V(T)^- \nu} \!\!\!&=&\!\!\!
\frac{d\Omega_{V}}{4 \pi}\ 
\frac{1}{\Gamma}\ 
\frac{1}{8 \pi}\ 
|V_{ud}|^2 \left( \frac{g_V}{m_V^2} \right)^2
G_{\rm F}^2 m_\tau^4 m_V^2 (1-\frac{m_V^2}{m_\tau^2}) 
\frac{2 m_V^2}{m_\tau^2}\ ,
\\
dR^{\tau^- \to V(T)^- \nu}_a \!\!\!&=&\!\!\!
dB^{\tau^- \to V(T)^- \nu}
\left(
\begin{array}{c}
- \sin \theta_V \cos \phi_V \\
- \sin \theta_V \sin \phi_V \\
- \cos \theta_V \\
\end{array}
\right)\ .
\end{eqnarray}
For leptonic decays,
after integrating over the phase space of the neutrinos,
the branching ratio and the spin dependence term are
given by
\begin{eqnarray}
 dB^{\tau^- \to l^- \bar{\nu} \nu} 
\!\!\!&=&\!\!\!
\frac{d\Omega_l}{4 \pi}\ dx\ 
\frac{1}{\Gamma}\ 
\frac{G_{\rm F}^2 m_\tau^5}{192 \pi^3}\ 
2 x^2 (3 - 2 x) \ ,\\
 dR^{\tau^- \to l^- \bar{\nu} \nu}_a
\!\!\!&=&\!\!\!
\frac{d\Omega_l}{4 \pi}\ dx\ 
\frac{1}{\Gamma}\ 
\frac{G_{\rm F}^2 m_\tau^5}{192 \pi^3}\ 
2 x^2 (1 - 2 x)
\left(
\begin{array}{c}
 \sin \theta_l \cos \phi_l \\
 \sin \theta_l \sin \phi_l \\
 \cos \theta_l \\
\end{array}
\right)
\ ,
\label{35}
\end{eqnarray}
where we neglect the masses of the leptons ($e$ and $\mu$) and
$x$ is the lepton energy normalized by the 
maximum energy $m_\tau/2$ 
i.e.\ $x=2 E_l/m_\tau$,
and ($\theta_l, \phi_l$) are the polar angles of the 
lepton in Frame 3.

Substituting these results into the formula in eq.(\ref{neo1}),
we obtain the differential cross sections
of each processes.
For example,
the differential cross section
of $e^+ e^- \to \tau^+ \tau^- \to \mu^+ \gamma + \pi^- \nu$ process
is given by
\begin{eqnarray}
&&\!\!\!\!\!\!\!\!\!\!
d\sigma (e^+ e^- \to \tau^+ \tau^- \to \mu^+ \gamma + \pi^- \nu)
\nonumber \\
\!\!\!&=&\!\!\!
\frac{\sigma (e^+e^- \to \tau^+ \tau^-)}
{\frac{4}{3} \left( 1 + \frac{2 m_\tau^2}{s}  \right)}\ 
B(\tau^- \to \pi^- \nu)\ 
B(\tau^+ \to \mu^+ \gamma)\ 
\frac{d\Omega_{\tau}}{4 \pi}\ 
\frac{d\Omega_{\gamma}}{4 \pi}\ 
\frac{d\Omega_{\pi}}{4 \pi}\ 
\nonumber \\
&& \!\!\!
\times 
\left[ {\rule[-3mm]{0mm}{10mm}\ } \right.
\left( 1+ \frac{4 m_\tau^2}{s} \right)
+ \left( 1- \frac{4 m_\tau^2}{s} \right) \cos^2 \theta_\tau
\nonumber \\
&&\ \ \ \
+ A_P 
\left(
\begin{array}{ccc}
 \sin \theta_\pi \cos \phi_\pi & \sin \theta_\pi \sin \phi_\pi 
& \cos \theta_\pi \\
\end{array}
\right)
\nonumber \\
&&\ \ \ \ 
\times
\left(
\begin{array}{ccc}
 \left( 1+ \frac{4 m_\tau^2}{s} \right) \sin^2 \theta_\tau & 0 
 & -\frac{2 m_\tau}{\sqrt{s}} \sin 2 \theta_\tau \\
 0 & \left( 1- \frac{4 m_\tau^2}{s} \right) \sin^2 \theta_\tau & 0 \\
 \frac{2 m_\tau}{\sqrt{s}} \sin 2 \theta_\tau & 0 
 & - \left( 1- \frac{4 m_\tau^2}{s} \right) - 
\left( 1+ \frac{4 m_\tau^2}{s} \right)
\cos^2 \theta_\tau \\
\end{array}
\right)
\nonumber \\
&&\ \ \ \ 
\times
\left(
\begin{array}{c}
 \sin \theta_\mu \cos \phi_\mu \\
 \sin \theta_\mu \sin \phi_\mu \\
 \cos \theta_\mu \\
\end{array}
\right)
\left. {\rule[-3mm]{0mm}{10mm}\ } \right]\ ,
\label{31}
\end{eqnarray}
where
\begin{eqnarray}
 \sigma (e^+ e^- \to \tau^+ \tau^-) =
\frac{4 \pi \alpha^2}{3 s}\ 
\sqrt{1-\frac{4 m_\tau^2}{s}}
\left(
1 + \frac{2 m_\tau^2}{s}
\right)
\end{eqnarray}
is the $\tau^+ \tau^-$ production cross section.
The branching ratio of $\tau^- \to \pi^- \nu$ and
$\tau^+ \to \mu^+ \gamma$ is given by
\begin{eqnarray}
 B (\tau^- \to \pi^- \nu) =
\frac{1}{\Gamma}\ 
\frac{1}{8 \pi}\ 
|V_{ud}|^2 f_\pi^2 G_{\rm F}^2 m_\tau^3\ ,
\end{eqnarray}
\begin{eqnarray}
 B (\tau^+ \to \mu^+ \gamma) =
\frac{1}{\Gamma}\ 
\frac{2}{\pi}\ 
G_{\rm F}^2 m_\tau^5
( |A_L|^2 + |A_R|^2 )\ ,
\end{eqnarray}
and the asymmetry parameter $A_P$ is defined as follows:
\begin{eqnarray}
 A_P \equiv \frac{|A_L|^2 - |A_R|^2}{|A_L|^2 + |A_R|^2 }\ .
\end{eqnarray}
We can see that
the measurement of
angular correlation of the pion and muon momentum
enables us to determine the parameter $A_P$,
so that we can obtain $|A_L|^2$ and $|A_R|^2$ separately.

A simpler expressions can be obtained
if we integrate over 
the angle $\theta_\tau$, $\phi_\tau$, $\phi_\pi$, and $\phi_\gamma$
in eq.(\ref{31}).
The differential cross section is given by
\begin{eqnarray}
&&\!\!\!\!
d\sigma (e^+ e^- \to \tau^+ \tau^- \to \mu^+ \gamma + \pi^- \nu)
\nonumber \\
&& = \sigma (e^+ e^- \to \tau^+ \tau^-) B(\tau^+ \to \mu^+ \gamma)
B (\tau^- \to \pi^- \nu) \frac{d \cos \theta_\mu}{2}
\frac{d \cos \theta_\pi}{2} \nonumber \\
&& \hspace*{1cm} \times
\left(
1 - \frac{s-2m_\tau^2}{s+2 m_\tau^2}\ A_P\ 
\cos \theta_\mu \cos \theta_\pi
\right) \ .
\label{41}
\end{eqnarray}
Notice that angular distribution in the
rest frames of $\tau^+$ and $\tau^-$
can be easily converted to the 
energy distribution in the
center of mass frame of the $e^+ e^-$ collision.
We obtain
\begin{eqnarray}
&&\!\!\!\!
d\sigma (e^+ e^- \to \tau^+ \tau^- \to \mu^+ \gamma + \pi^- \nu)
\nonumber \\
&& = \sigma (e^+ e^- \to \tau^+ \tau^-) B(\tau^+ \to \mu^+ \gamma)
B (\tau^- \to \pi^- \nu) \
\frac{s}{s-4 m_\tau^2}\ dz_\mu dz_\pi
\nonumber \\
&& \hspace*{1cm} \times
\left(
1 - \frac{s(s-2m_\tau^2)}{(s-4 m_\tau^2)(s+2 m_\tau^2)}\ A_P\ 
(2 z_\mu -1)(2 z_\pi -1)
\right) \ ,
\label{42}
\end{eqnarray}
where
$z_\mu=E_\mu/E_\tau$ ($z_\pi = E_\pi / E_\tau$), and 
$E_\mu$, $E_\pi$, and $E_\tau = \sqrt{s}/2$ are the 
energies of the muon, pion, and $\tau$
in the center of mass frame, respectively.

The angular (or energy) distributions in eq.(\ref{41}) (eq.(\ref{42}))
can be understood as follows.
Because of the helicity conservation of the $\tau^+ \tau^-$ production
process, 
the helicities of $\tau^+$ and $\tau^-$ are correlated,
namely $\tau^+_L \tau^-_R$ or $\tau^+_R \tau^-_L$
is produced.
This means that two $\tau$ spins are parallel
in the limit of $\sqrt{s} \gg m_\tau$.
In the decay process,
the $\pi^-$ tends to be emitted to the
spin direction of $\tau^-$ for $\tau^- \to \pi^- \nu$,
because of the $V-A$ interaction.
On the other hand,
for $\tau^+ \to \mu^+ \gamma$ decay,
the muon tends to be emitted to the same
direction of the $\tau^+$ spin if $A_P>0$.
Therefore the differential branching ratio is
enhanced (suppressed) if the sign of 
$\cos \theta_\mu \cos \theta_\pi$ is negative (positive).
In other words,
pion and muon energies in the center of mass frame
of the $e^+ e^-$ collision have a negative correlation
if $A_P>0$.
If $A_P<0$,
we have an opposite correlation.

\begin{figure}
\psfrag{u}{\it \Large u}
\psfrag{v}{\it \Large v}
\psfrag{+1}{\Huge +1}
\psfrag{-1}{\Huge --1}
\psfrag{0}{0}
\begin{center}
\includegraphics[width=10cm]{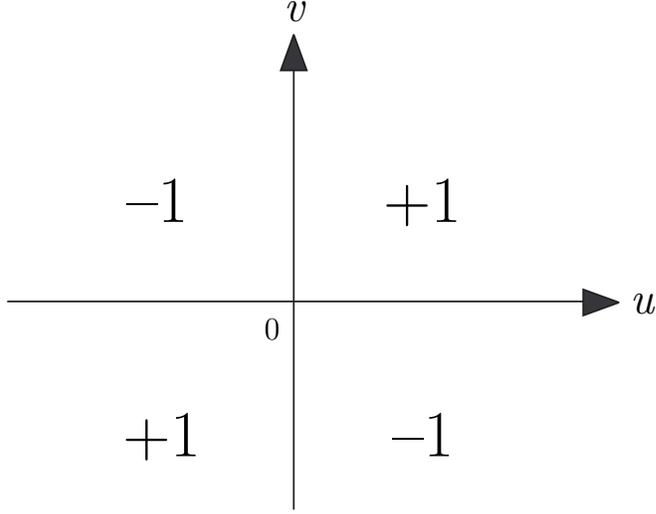} 
\end{center}
\caption{The weight function $w(u,v)$.}
\label{fig2}
\end{figure}

\begin{figure}
\begin{center}
\includegraphics[width=13cm]{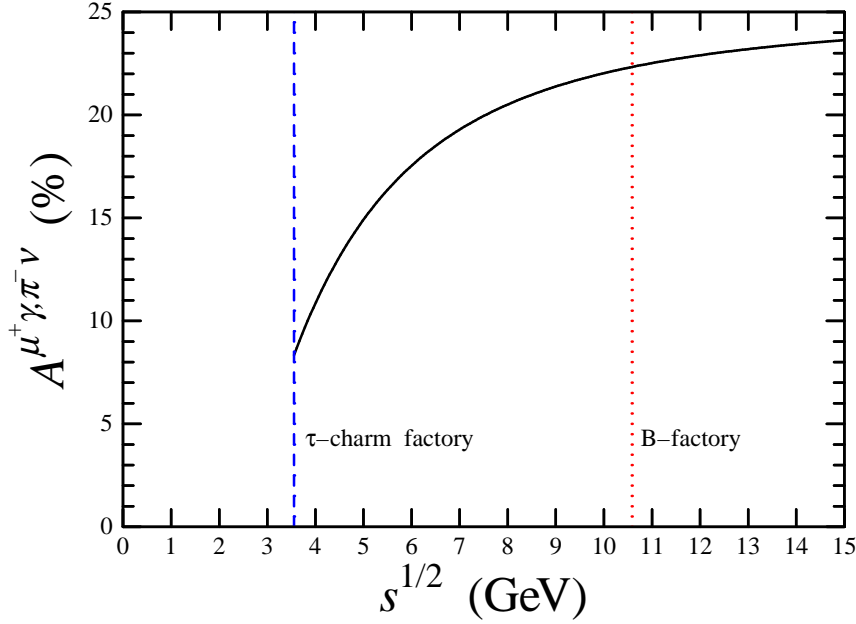} 
\end{center}
\caption{The observable asymmetry $A^{\mu^+ \gamma, \pi^- \nu}$ 
vs $\sqrt{s}$ for
$A_P=-1$.
The dashed line represents the $\sqrt{s}$ of 
$\tau$-charm factory and the dotted line represents that of
B-factory.
}
\label{figasym}
\end{figure}

We can define an asymmetry 
$A^{\mu^+ \gamma, \pi^- \nu}$
by the following asymmetric integrations
\begin{eqnarray}
 A^{\mu^+ \gamma, \pi^- \nu} 
\!\!\!&=&\!\!\! 
\frac{
\int d \cos \theta_\mu \ d \cos \theta_\pi \ 
w(\cos \theta_\mu, \cos \theta_\pi) \ 
\frac{d^2 \sigma}{d \cos \theta_\mu d\cos \theta_\pi}\ }{
\sigma (e^+ e^- \to \tau^+ \tau^-) B(\tau^+ \to \mu^+ \gamma)
B (\tau^- \to \pi^- \nu)  }
\nonumber \\
&=&\!\!\!
\frac{N^{++}+N^{--}-N^{+-}-N^{-+}}{N^{++}+N^{--}+N^{+-}+N^{-+}}
\ ,
\label{43}
\end{eqnarray}
where the weight function $w(u,v)$ is defined by
\begin{eqnarray}
 w(u,v) = \frac{uv}{|uv|}\ ,
\end{eqnarray}
and shown in Fig.\ref{fig2}.
In the second line,
$N^{\pm \pm}$ are the event numbers
where the first $\pm$ represent the sign of $\cos \theta_\mu$
and the second one is that of $\cos \theta_\pi$, respectively.
$A^{\mu^+ \gamma, \pi^- \nu}$ is related to the 
parameter $A_P$ by 
\begin{eqnarray}
 A^{\mu^+ \gamma, \pi^- \nu} =
- \frac{ s - 2 m_\tau^2 }{ 4 ( s + 2 m_\tau^2 ) } A_P \ .
\end{eqnarray}
In Fig.\ref{figasym},
$\sqrt{s}$ dependence of $A^{\mu^+ \gamma, \pi^- \gamma}$
is shown for $A_P = -1$.
We can see that the asymmetry is already close to the maximal value
at the B-factory energy.

It is straightforward to extend the above formula
to other cases.
We only present here formulae corresponding to eq.(\ref{41}) 
for different decay modes of $\tau^-$.
\begin{eqnarray}
&&\!\!\!\!
d\sigma (e^+ e^- \to \tau^+ \tau^- \to \mu^+ \gamma + V^- \nu)
\nonumber \\
&& = \sigma (e^+ e^- \to \tau^+ \tau^-) B(\tau^+ \to \mu^+ \gamma)
B (\tau^- \to V^- \nu) \frac{d \cos \theta_\mu}{2}
\frac{d \cos \theta_V}{2} \nonumber \\
&& \hspace*{1cm} \times
\left(
1 \pm \frac{s-2m_\tau^2}{s+2 m_\tau^2}\ A_P\ 
\cos \theta_\mu \cos \theta_V
\right) \ ,
\label{44}
\end{eqnarray}
where $+$ corresponds to the vector mesons
with transverse polarization $V=\rho (T) $, $a_1 (T)$
and $-$ corresponds to those with longitudinal polarization
$V= \rho (L)$, $a_1 (L)$.
For leptonic decay, we obtain
\begin{eqnarray}
 &&\!\!\!\!
d\sigma (e^+ e^- \to \tau^+ \tau^- \to \mu^+ \gamma + l^- \bar{\nu} \nu)
\nonumber \\
&& = \sigma (e^+ e^- \to \tau^+ \tau^-) B(\tau^+ \to \mu^+ \gamma)
B (\tau^- \to l^- \bar{\nu} \nu) \frac{d \cos \theta_\mu}{2}
\frac{d \cos \theta_l}{2} \ dx \ 2x^2\nonumber \\
&& \hspace*{1cm} \times
\left \{
3-2x - \frac{s-2m_\tau^2}{s+2 m_\tau^2} (1-2x) A_P
\cos \theta_\mu \cos \theta_l
\right \}\ .
\label{45}
\end{eqnarray}
The measurement of the polarization of the
vector mesons can be done by the analysis of 
the distribution of the two (or three) pions 
from the $\rho$ ($a_1$) meson decay \cite{Bullock:1993yt}.

In the case of $\tau^-$ decays into $\mu^- \gamma$ and
$\tau^+$ decays via $V-A$ interaction,
the $dR_a^{\tau^- \to f_A}$ and $dR_b^{\tau^+ \to f_B}$ acquire
extra minus signs. 
For example,
\begin{eqnarray}
 dR^{\tau^- \to \mu^- \gamma}_a =
- \frac{d \Omega^\prime_\mu}{4 \pi}\ 
\frac{1}{\Gamma}\ \frac{2}{\pi}\ 
G_{\rm F}^2 m_\tau^5\ 
(|A_L|^2 - |A_R|^2)\ 
\left(
\begin{array}{c}
 \sin \theta^\prime_\mu \cos \phi^\prime_\mu \\
 \sin \theta^\prime_\mu \sin \phi^\prime_\mu \\
 \cos \theta^\prime_\mu \\
\end{array}
\right)\ ,
\end{eqnarray}
\begin{eqnarray}
 dR^{\tau^+ \to \pi^+ \bar{\nu}}_b =
- \frac{d\Omega^\prime_\pi}{4 \pi}\ 
\frac{1}{\Gamma}\
\frac{1}{8 \pi}\
|V_{ud}|^2 f_\pi^2 G_{\rm F}^2 m_\tau^3
\left(
\begin{array}{c}
 \sin \theta^\prime_\pi \cos \phi^\prime_\pi \\
 \sin \theta^\prime_\pi \sin \phi^\prime_\pi \\
 \cos \theta^\prime_\pi \\
\end{array}
\right)
\ ,
\end{eqnarray}
where ($\theta^\prime_\mu$, $\phi^\prime_\mu$) 
(($\theta^\prime_\pi$, $\phi^\prime_\pi$))
are the polar angle of the muon (pion) momentum
in Frame 3 (Frame 2).
The formula in eq.(\ref{41})
can be applied to the $\tau^- \to \mu^- \gamma$ case
by the replacement of 
$(\theta_\mu, \theta_\pi)$
by
$(\theta_\mu^\prime, \theta_\pi^\prime)$,
and therefore same angular and energy correlation
holds as in the $\tau^+ \to \mu^+ \gamma$ case.
In a similar way, we can obtain the formulae
corresponds to eqs.(\ref{44}) and (\ref{45})
for the $\tau^- \to \mu^- \gamma$ case by
replacement of 
$(\theta_V, \theta_l)$ by 
$(\theta_V^\prime, \theta_l^\prime)$,
where $\theta^\prime_V$ ($\theta^\prime_l$) is
the angle between the vector meson (lepton) momentum
and $\tau^+$ direction in Frame 2.

\section{P and T asymmetries in LFV three body $\tau$ decays}

\begin{figure}
\psfrag{mu^+}{$\mu^+$}
\psfrag{mu^-}{$\mu^-$}
\psfrag{x_4}{$x_4$}
\psfrag{y_4}{$y_4$}
\psfrag{z_4}{$z_4$}
\begin{center}
\includegraphics[width=10cm]{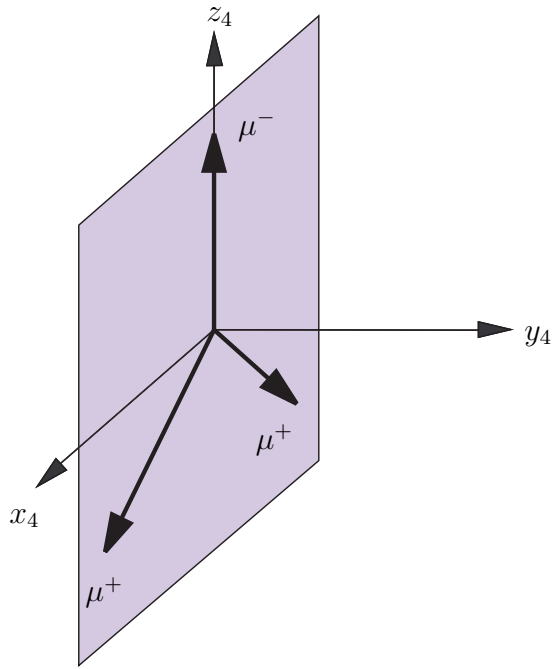} 
\vspace*{1cm}
\end{center}
\caption{The coordinate system in the $\tau \to 3 \mu$ calculation.}
\label{fig3muplane}
\end{figure}
\begin{figure}
\psfrag{th}{$\theta$}
\psfrag{phi}{$\phi$}
\psfrag{psi}{$\psi$}
\psfrag{x_4}{$x_4$}
\psfrag{y_4}{$y_4$}
\psfrag{z_4}{$z_4$}
\psfrag{x_2}{$x_2$}
\psfrag{y_2}{$y_2$}
\psfrag{z_2}{$z_2$}
\begin{center}
\includegraphics[width=10cm]{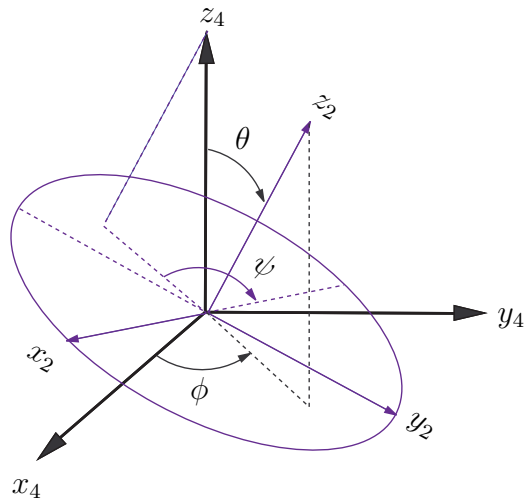} 
\vspace*{1cm}
\end{center}
\caption{The relation between Frame 2 and Frame 4.}
\label{fig3mu}
\end{figure}

In this section,
we consider LFV three body decays
i.e.\ $\tau \to 3\mu$, $\tau \to 3e$, $\tau \to \mu ee$, and
$\tau \to e \mu \mu$.
Within the approximation that the 
muon and electron masses are neglected,
$\tau \to 3 \mu$ and $\tau \to 3e$
(or $\tau \to \mu ee$ and $\tau \to e \mu \mu$)
give the same formula,
so that we only consider $\tau \to 3 \mu$ and 
$\tau \to \mu ee$ processes.
In these processes, we can define
the P odd as well as T odd asymmetries
of $\tau$ decays.

For $\tau^+ \to \mu^+ \mu^+ \mu^-$ decay,
the effective Lagrangian is given by
\begin{eqnarray}
  {\cal L} &=& -\frac{4 G_{\rm F}}{\sqrt{2}}
\left \{
m_\tau A_R \bar{\tau} \sigma^{\mu \nu} P_L \mu F_{\mu \nu}
+ m_\tau A_L \bar{\tau} \sigma^{\mu \nu} P_R \mu F_{\mu \nu} 
\right. \nonumber \\
&& \hspace*{1.5cm}
+ g_1 (\bar{\tau} P_L \mu)(\bar{\mu} P_L \mu) 
+ g_2 (\bar{\tau} P_R \mu)(\bar{\mu} P_R \mu) \nonumber \\
&& \hspace*{1.5cm}
+ g_3 (\bar{\tau} \gamma^\mu P_R \mu)(\bar{\mu} \gamma_\mu P_R \mu) 
+ g_4 (\bar{\tau} \gamma^\mu P_L \mu)(\bar{\mu} \gamma_\mu P_L \mu) 
\nonumber \\
&& \hspace*{1.5cm}
\left.
+ g_5 (\bar{\tau} \gamma^\mu P_R \mu)(\bar{\mu} \gamma_\mu P_L \mu) 
+ g_6 (\bar{\tau} \gamma^\mu P_L \mu)(\bar{\mu} \gamma_\mu P_R \mu) 
+ {\rm h.c.}\right \}\ .
\label{39}
\end{eqnarray}
With this Lagrangian in eq.(\ref{39}),
we can calculate the differential branching 
ratio $dB^{\tau^+ \to 3\mu}$
and the spin dependence term $dR^{\tau^+ \to 3\mu}_b$
in eq.(\ref{neo1}).
In order to calculate these quantities
we first define the 
Lorentz frame (Frame 4)
for the three body decays \cite{Okada:2000zk}.
Frame 4 is the rest frame of $\tau^+$ and
we take $z$-direction to be $\mu^-$ momentum direction, 
and $xz$-plane to be the decay plane.
The $x$-direction is determined 
so that the $x$-component of the momentum for 
the $\mu^+$ with larger energy is positive.
The coordinate system are shown in Fig.\ref{fig3muplane}.
Any four vector in
Frame 4 is
related to that in Frame 2
by the Euler rotation with three angles 
($\theta$, $\phi$, $\psi$) as follows (Fig.\ref{fig3mu}):
\begin{eqnarray}
 \xi_4^\mu \!\!\!&=&\!\!\!
\nonumber \\
&&\!\!\!\!\!\!\!\!
\left(
\begin{array}{cccc}
 1 & 0 & 0 & 0 \\
 0 & \cos \phi & -\sin \phi & 0 \\
 0 & \sin \phi & \cos \phi & 0 \\
 0 & 0 & 0 & 1 \\
\end{array}
\right)
\left(
\begin{array}{cccc}
 1 & 0 & 0 & 0 \\
 0 & \cos \theta & 0 & \sin \theta \\
 0 & 0 & 1 & 0 \\
 0 & -\sin \theta & 0 & \cos \theta \\
\end{array}
\right)
\left(
\begin{array}{cccc}
 1 & 0 & 0 & 0 \\
 0 & \cos \psi & -\sin \psi & 0 \\
 0 & \sin \psi & \cos \psi & 0 \\
 0 & 0 & 0 & 1 \\
\end{array}
\right)
\xi_2^\mu\ ,
\nonumber \\
\label{40}
\end{eqnarray}
where 
\begin{eqnarray}
 0 \leq \theta \leq \pi\ \ \ ,\ \ \ 
 0 \leq \phi \leq 2 \pi\ \ \ ,\ \ \ 
 0 \leq \psi \leq 2 \pi\ .
\end{eqnarray}
We also define the energy variables $x_1 = 2E_1/m_\tau$ 
and $x_2 = 2 E_2/ m_\tau$
where $E_1$ ($E_2$) is the energy of $\mu^+$
with a larger (smaller) energy in the rest frame of $\tau^+$.

With these angles and energy variables,
the branching ratio and spin dependence term
can be expressed as follows:
\begin{eqnarray}
 dB^{\tau^+ \to 3\mu} =
\frac{1}{\Gamma}\ 
\frac{m_\tau^5 G_{\rm F}^2}{256 \pi^5}\ 
dx_1 \ dx_2 \ d \cos \theta\  d\phi \ d\psi \  X\ ,
\end{eqnarray}
\begin{eqnarray}
 dR^{\tau^+ \to 3\mu}_b \!\!\!&=&\!\!\!
\frac{1}{\Gamma}\ 
\frac{m_\tau^5 G_{\rm F}^2}{256 \pi^5}\ 
dx_1 \ dx_2 \ d \cos \theta\  d\phi \ d\psi \  
\nonumber \\
&& \times
\left(
\begin{array}{c}
 -Y s_\theta\ c_\psi
+ Z (c_\theta c_\phi c_\psi - s_\phi s_\psi)
+ W (c_\theta s_\phi c_\psi + c_\phi s_\psi)\\
 Y s_\theta\ s_\psi
+ Z (-c_\theta c_\phi s_\psi - s_\phi c_\psi)
+ W (-c_\theta s_\phi s_\psi + c_\phi c_\psi)\\
 Y c_\theta + Z s_\theta c_\phi + W s_\theta s_\phi  \\
\end{array}
\right)\ ,
\label{54}
\end{eqnarray}
where $s_\theta$ ($s_\phi$, $s_\psi$) and 
$c_\theta$ ($c_\phi$, $c_\psi$) represent
$\sin \theta$ ($\sin \phi$, $\sin \psi$) and 
$\cos \theta$ ($\cos \phi$, $\cos \psi$), respectively.
The functions $X$, $Y$, $Z$, and $W$ are defined as follows:
\begin{eqnarray}
 X&=&\left(
\frac{|g_1|^2}{16} + \frac{|g_2|^2}{16} +|g_3|^2 +|g_4|^2
\right) \alpha_1 (x_1,x_2)
+\left(
|g_5|^2 + |g_6|^2
\right) \alpha_2 (x_1,x_2) \nonumber \\
&&+ \left(
|e A_R|^2 + |e A_L|^2
\right) \alpha_3 (x_1,x_2)
- {\rm Re} (e A_R g_4^* + e A_L g_3^*)\  \alpha_4 (x_1,x_2)
\nonumber \\
&& -{\rm Re} (e A_R g_6^* + e A_L g_5^*) \ \alpha_5 (x_1,x_2) \ ,
\end{eqnarray}
\begin{eqnarray}
 Y &=&
\left(
\frac{|g_1|^2}{16} - \frac{|g_2|^2}{16} +|g_3|^2 -|g_4|^2
\right) \alpha_1 (x_1,x_2)
+ {\rm Re} (e A_R g_4^* - eA_L g_3^*)\ \alpha_4 (x_1,x_2)
\nonumber \\
&&
- {\rm Re} (e A_R g_6^* - e A_L g_5^*) \ \alpha_5 (x_1, x_2)
+\left(
|g_5|^2-|g_6|^2
\right) \beta_1 (x_1,x_2)
\nonumber \\
&&
+ \left(
|e A_R|^2 - |e A_L|^2
\right) \beta_2 (x_1,x_2)\ ,
\end{eqnarray}
\begin{eqnarray}
 Z &=&
\left(
|g_5|^2 - |g_6|^2
\right) \gamma_1 (x_1,x_2)
+ \left(
|e A_R|^2 - |e A_L|^2
\right) \gamma_2 (x_1,x_2)
\nonumber \\
&&
- {\rm Re}
\left(
e A_R g_4^* - e A_L g_3^*
\right) \gamma_3 (x_1,x_2)
+ {\rm Re}
\left(
e A_R g_6^* - e A_L g_5^*
\right) \gamma_4 (x_1,x_2)\ ,
\end{eqnarray}
\begin{eqnarray}
 W&=&
- {\rm Im}
\left(
e A_R g_4^* + e A_L g_3^*
\right) \gamma_3 (x_1,x_2)
+ {\rm Im}
\left(
e A_R g_6^* + e A_L g_5^*
\right) \gamma_4 (x_1,x_2)\ ,
\end{eqnarray}
where $e$ ($>0$) is the positron charge
and functions $\alpha_{1-5}$, $\beta_{1-2}$,
and $\gamma_{1-4}$ are given in Appendix \ref{functions}.
Notice that the $Y$ and $Z$ terms represent P odd
quantities with respect to the $\tau^+$ spin in the rest
frame of $\tau^+$ and the $W$ term represents a T odd quantity.
These are the same as P and T odd terms considered
in the differential decay width of 
$\mu^+ \to e^+ e^+ e^-$ \cite{Okada:2000zk}.

The differential cross section is 
obtained by substituting this into eq.(\ref{neo1}).
In the case that the opposite side $\tau$ 
decays into $\pi^- \nu$,
we obtain after integrating over 
$\phi_\pi$, $\psi$, $\theta_\tau$, and $\phi_\tau$
\begin{eqnarray}
 &&d\sigma (e^+ e^- \to \tau^+ \tau^- \to \mu^+ \mu^+ \mu^- + \pi^- \nu)
\nonumber \\
&& = \sigma (e^+ e^- \to \tau^+ \tau^-) 
B (\tau^- \to \pi^- \nu) \left(
\frac{m_\tau^5 G_{\rm F}^2}{128 \pi^4}/\Gamma
\right)
\frac{d \cos \theta_\pi}{2}\ dx_1 \ dx_2\ d\cos \theta \ d\phi
\nonumber \\
&& \hspace*{1cm} \times
\left[
X - \frac{s-2m_\tau^2}{s+2 m_\tau^2} 
\left \{ 
Y \cos \theta +
Z \sin \theta \cos \phi +
W \sin \theta \sin \phi
\right \}
\cos \theta_\pi 
\right] \ .
\end{eqnarray}
The terms $X$, $Y$, $Z$, and $W$ can be extracted 
by the following (asymmetric) integrations.
\begin{eqnarray}
&&
\int d \cos \theta \ d \phi\  d \cos \theta_\pi\
\frac{d^5 \sigma}{dx_1 dx_2 d \cos \theta\ d\phi\ d \cos \theta_\pi} 
\nonumber \\
&&\times 
\left \{
\sigma (e^+ e^- \to \tau^+ \tau^-) 
B (\tau^- \to \pi^- \nu)
\left(
\frac{m_\tau^5 G_{\rm F}^2}{32 \pi^3}/\Gamma
\right)
\right \}^{-1} = X
\ ,
\label{59}
\end{eqnarray}
\begin{eqnarray}
&&
\int d \cos \theta \ d \cos \theta_\pi\
w (\cos \theta, \cos \theta_\pi)\
\frac{d^4 \sigma}{dx_1 dx_2\  d \cos \theta\ d \cos \theta_\pi} 
\nonumber \\
&&\times 
\left \{
\sigma (e^+ e^- \to \tau^+ \tau^-) 
B (\tau^- \to \pi^- \nu)
\left(
\frac{m_\tau^5 G_{\rm F}^2}{32 \pi^3}/\Gamma
\right)
\right \}^{-1}
= - \frac{(s-2 m_\tau^2)}{4(s+2m_\tau^2)} Y
\ ,
\end{eqnarray}
\begin{eqnarray}
&&
\int d \phi \ d \cos \theta_\pi\
w (\cos \phi, \cos \theta_\pi)\
\frac{d^4 \sigma}{dx_1 dx_2\  d \phi\ d \cos \theta_\pi} 
\nonumber \\
&&\times 
\left \{
\sigma (e^+ e^- \to \tau^+ \tau^-) 
B (\tau^- \to \pi^- \nu)
\left(
\frac{m_\tau^5 G_{\rm F}^2}{32 \pi^3}/\Gamma
\right)
\right \}^{-1}
= - \frac{(s-2 m_\tau^2)}{4(s+2m_\tau^2)} Z
\ ,
\end{eqnarray}
\begin{eqnarray}
&& 
\int d \phi \ d \cos \theta_\pi\
w (\sin \phi, \cos \theta_\pi)\
\frac{d^4 \sigma}{dx_1 dx_2\  d \phi\ d \cos \theta_\pi} 
\nonumber \\
&&\times 
\left \{
\sigma (e^+ e^- \to \tau^+ \tau^-) 
B (\tau^- \to \pi^- \nu)
\left(
\frac{m_\tau^5 G_{\rm F}^2}{32 \pi^3}/\Gamma
\right)
\right \}^{-1}
= - \frac{(s-2 m_\tau^2)}{4(s+2m_\tau^2)} W
\ .
\label{62}
\end{eqnarray}
Notice that
the function $W$ represents CP violating LFV interaction.
We can see that this
is induced by the relative phase between
the photon-penguin coupling constants ($A_L$ and $A_R$)
and the four-fermion coupling constants ($g_3 - g_6$).

A similar formula can be obtained
for the $\tau^+ \to \mu^+ e^+ e^-$ decay.
The effective Lagrangian for the $\tau^+ \to \mu^+ e^+ e^-$
is given by
\begin{eqnarray}
 {\cal L} &=& -\frac{4 G_{\rm F}}{\sqrt{2}}
\left \{
m_\tau A_R \bar{\tau} \sigma^{\mu \nu} P_L \mu F_{\mu \nu}
+ m_\tau A_L \bar{\tau} \sigma^{\mu \nu} P_R \mu F_{\mu \nu} 
\right. \nonumber \\
&& \hspace*{1.5cm}
+ \lambda_1 (\bar{\tau} P_L \mu)(\bar{e} P_L e) 
+ \lambda_2 (\bar{\tau} P_L \mu)(\bar{e} P_R e) \nonumber \\
&& \hspace*{1.5cm}
+ \lambda_3 (\bar{\tau} P_R \mu)(\bar{e} P_L e) 
+ \lambda_4 (\bar{\tau} P_R \mu)(\bar{e} P_R e) \nonumber \\
&& \hspace*{1.5cm}
+ \lambda_5 (\bar{\tau} \gamma^\mu P_L \mu)(\bar{e} \gamma_\mu P_L e) 
+ \lambda_6 (\bar{\tau} \gamma^\mu P_L \mu)(\bar{e} \gamma_\mu P_R e) 
\nonumber \\
&& \hspace*{1.5cm}
+ \lambda_7 (\bar{\tau} \gamma^\mu P_R \mu)(\bar{e} \gamma_\mu P_L e) 
+ \lambda_8 (\bar{\tau} \gamma^\mu P_R \mu)(\bar{e} \gamma_\mu P_R e) 
\nonumber \\
&& \hspace*{1.5cm}
\left.
+ \lambda_9 (\bar{\tau} \sigma^{\mu \nu} P_L \mu)
(\bar{e} \sigma_{\mu \nu} e) 
+ \lambda_{10} (\bar{\tau} \sigma^{\mu \nu} P_R \mu)
(\bar{e} \sigma_{\mu \nu} e) 
+ {\rm h.c.}\right \}\ .
\end{eqnarray}
In this calculation,
we define Frame 4$^\prime$ which is almost the same as
Frame 4 in the $\tau^+ \to \mu^+ \mu^+ \mu^-$ case.
The definition is obtained 
by the replacement of $\mu^-$ of $\tau^+ \to \mu^+ \mu^+ \mu^-$
by $e^-$ of $\tau^+ \to \mu^+ e^+ e^-$,
the $\mu^+$ with a larger energy
of $\tau^+ \to \mu^+ \mu^+ \mu^-$
by $\mu^+$ of $\tau^+ \to \mu^+ e^+ e^-$,
and
$\mu^+$ with a smaller energy of 
$\tau^+ \to \mu^+ \mu^+ \mu^-$
by $e^+$ of $\tau^+ \to \mu^+ e^+ e^-$.
If we take 
the definition of ($\theta$, $\phi$, $\psi$)
in such a way that
the same relation is satisfied as in eq.(\ref{40}),
the branching ratio and the spin dependence term
are given by
\begin{eqnarray}
 dB^{\tau^+ \to \mu^+ e^+ e^-} =
\frac{1}{\Gamma}\ 
\frac{m_\tau^5 G_{\rm F}^2}{256 \pi^5}\ 
dx_1 \ dx_2 \ d \cos \theta\  d\phi \ d\psi \  X^\prime\ ,
\end{eqnarray}
\begin{eqnarray}
 dR^{\tau^+ \to \mu^+ e^+ e^-}_b \!\!\!&=&\!\!\!
\frac{1}{\Gamma}\ 
\frac{m_\tau^5 G_{\rm F}^2}{256 \pi^5}\ 
dx_1 \ dx_2 \ d \cos \theta\  d\phi \ d\psi \  
\nonumber \\
&& \times
\left(
\begin{array}{c}
 -Y^\prime s_\theta\ c_\psi
+ Z^\prime (c_\theta c_\phi c_\psi - s_\phi s_\psi)
+ W^\prime (c_\theta s_\phi c_\psi + c_\phi s_\psi)\\
 Y^\prime s_\theta\ s_\psi
+ Z^\prime (-c_\theta c_\phi s_\psi - s_\phi c_\psi)
+ W^\prime (-c_\theta s_\phi s_\psi + c_\phi c_\psi)\\
 Y^\prime c_\theta + Z^\prime s_\theta c_\phi + W^\prime s_\theta s_\phi  \\
\end{array}
\right) , \nonumber \\
\label{neo66}
\end{eqnarray}
where the functions $X^\prime$, $Y^\prime$, $Z^\prime$, and $W^\prime$
are given by
\begin{eqnarray}
  X^\prime &=&
\left(
|e A_R|^2 + |e A_L|^2
\right) A_1 (x_1,x_2)
+ {\rm Re}
\left(
e A_R \lambda_5^* + e A_L \lambda_8^*
\right) A_2 (x_1,x_2) \nonumber \\
&& + {\rm Re}
\left(
eA_R \lambda_6^* + e A_L \lambda_7^*
\right) A_3 (x_2)
+ \left(
|\lambda_1|^2 + |\lambda_2|^2 + |\lambda_3|^2 + |\lambda_4|^2
\right) A_4 (x_1) \nonumber \\
&& + \left(
|\lambda_5|^2 + |\lambda_8|^2
\right) A_5 (x_1, x_2)
+ \left(
|\lambda_6|^2 + |\lambda_7|^2
\right) A_6 (x_2) \nonumber \\
&& + \left(
|\lambda_9|^2 + |\lambda_{10}|^2
\right) A_7 (x_1,x_2) 
+ {\rm Re} \left(
\lambda_1 \lambda_9^* + \lambda_4 \lambda_{10}^*
\right) A_8 (x_1,x_2)
\ ,
\end{eqnarray}
\begin{eqnarray}
 Y^\prime &=&
-{\rm Re} \left(
e A_R \lambda_5^* - e A_L \lambda_8^*
\right) A_2 (x_1,x_2)
+{\rm Re} \left(
e A_R \lambda_6^* - e A_L \lambda_7^*
\right) A_3 (x_1,x_2) \nonumber \\
&& -\left(
|\lambda_5|^2-|\lambda_8|^2
\right) A_5 (x_1,x_2)
+ \left(
|e A_R|^2 - |e A_L|^2
\right) B_1 (x_1,x_2) \nonumber \\
&& + \left(
|\lambda_1|^2 + |\lambda_2|^2 - |\lambda_3|^2 - |\lambda_4|^2
\right) B_2 (x_1,x_2)
+\left(
|\lambda_6|^2- |\lambda_7|^2
\right) B_3 (x_1,x_2) \nonumber \\
&& +\left(
|\lambda_9|^2 - |\lambda_{10}|^2
\right) B_4 (x_1,x_2)
+ {\rm Re} \left(
\lambda_1 \lambda_9^* - \lambda_4 \lambda_{10}^*
\right) B_5 (x_1,x_2)
\ ,
\end{eqnarray}
\begin{eqnarray}
 Z^\prime &=&
\left(
|e A_R|^2 - |e A_L|^2
\right) C_1 (x_1,x_2)
+ {\rm Re} \left(
e A_R \lambda_5^* - e A_L \lambda_8^*
\right) C_2 (x_1,x_2) \nonumber \\
&& +{\rm Re} \left(
e A_R \lambda_6^* - e A_L \lambda_7^*
\right) C_3 (x_1,x_2)
+ \left(
|\lambda_1|^2 + |\lambda_2|^2 - |\lambda_3|^2 - |\lambda_4|^2
\right) C_4 (x_1,x_2) \nonumber \\
&& + \left \{
|\lambda_6|^2 - |\lambda_7|^2
+ {\rm Re} \left(
-2 \lambda_1 \lambda_9^* +2 \lambda_4 \lambda_{10}^*
\right)
\right \} C_5 (x_1,x_2) 
\nonumber \\
&&
+ \left(
|\lambda_9|^2 -|\lambda_{10}|^2
\right) C_6 (x_1,x_2)
\ ,
\end{eqnarray}
\begin{eqnarray}
 W^\prime &=&
{\rm Im} \left(
e A_R \lambda_5^* + e A_L \lambda_8^*
\right) C_2 (x_1,x_2)
+{\rm Im} \left(
e A_R \lambda_6^* + e A_L \lambda_7^*
\right) C_3 (x_1,x_2) \nonumber \\
&&
+{\rm Im} \left(
\lambda_1 \lambda_{9}^* + \lambda_4 \lambda_{10}^*
\right) C_7 (x_1,x_2)
\ .
\end{eqnarray}
The functions $A_{1-7},B_{1-4},C_{1-6}$ are given 
in Appendix \ref{functions}.
The $X^\prime$, $Y^\prime$, $Z^\prime$, and $W^\prime$
can be extracted 
in the same way as in eqs.(\ref{59})--(\ref{62}).

Next we consider the decay mode of 
$\tau^+ \to \mu^- e^+ e^+$.
This case is different from above
in the point that
both $\tau \to e$ and $\mu \to e$ 
transitions are necessary.
The effective Lagrangian 
for this process
is given by
\begin{eqnarray}
  {\cal L} &=& -\frac{4 G_{\rm F}}{\sqrt{2}}
\left \{
 g^\prime_1 (\bar{\tau} P_L e)(\bar{\mu} P_L e) 
+ g^\prime_2 (\bar{\tau} P_R e)(\bar{\mu} P_R e) \right. \nonumber \\
&& \hspace*{1.5cm}
+ g^\prime_3 (\bar{\tau} \gamma^\mu P_R e)(\bar{\mu} \gamma_\mu P_R e) 
+ g^\prime_4 (\bar{\tau} \gamma^\mu P_L e)(\bar{\mu} \gamma_\mu P_L e) 
\nonumber \\
&& \hspace*{1.5cm}
\left.
+ g^\prime_5 (\bar{\tau} \gamma^\mu P_R e)(\bar{\mu} \gamma_\mu P_L e) 
+ g^\prime_6 (\bar{\tau} \gamma^\mu P_L e)(\bar{\mu} \gamma_\mu P_R e) 
+ {\rm h.c.}\right \}\ .
\end{eqnarray}
If we take a coordinate system
similar to Frame 4,
in which the larger (smaller) energy $\mu^+$ is replaced by
the larger (smaller) energy $e^+$,
$dB^{\tau^+ \to \mu^- e^+ e^+}$ and 
$dR_b^{\tau^+ \to \mu^- e^+ e^+}$ are given by
\begin{eqnarray}
 dB^{\tau^+ \to \mu^- e^+ e^+} =
\frac{1}{\Gamma}\ 
\frac{m_\tau^5 G_{\rm F}^2}{256 \pi^5}\ 
dx_1 \ dx_2 \ d \cos \theta\  d\phi \ d\psi \  X^{\prime \prime}\ ,
\end{eqnarray}
\begin{eqnarray}
 dR^{\tau^+ \to \mu^+ e^+ e^-}_b \!\!\!&=&\!\!\!
\frac{1}{\Gamma}\ 
\frac{m_\tau^5 G_{\rm F}^2}{256 \pi^5}\ 
dx_1 \ dx_2 \ d \cos \theta\  d\phi \ d\psi \  
\nonumber \\
&& \times
\left(
\begin{array}{c}
 -Y^{\prime \prime} s_\theta\ c_\psi
+ Z^{\prime \prime} (c_\theta c_\phi c_\psi - s_\phi s_\psi) \\
 Y^{\prime \prime} s_\theta\ s_\psi
+ Z^{\prime \prime} (-c_\theta c_\phi s_\psi - s_\phi c_\psi)
\\
 Y^{\prime \prime} c_\theta + Z^{\prime \prime} s_\theta c_\phi 
\\
\end{array}
\right)\ ,
\label{73}
\end{eqnarray}
where functions $X^{\prime \prime}$, $Y^{\prime \prime}$,
and $Z^{\prime \prime}$ are given by
\begin{eqnarray}
 X^{\prime \prime}&=&\left(
\frac{|g^\prime_1|^2}{16} + \frac{|g^\prime_2|^2}{16} 
+|g^\prime_3|^2 +|g^\prime_4|^2
\right) \alpha_1 (x_1,x_2)
+\left(
|g^\prime_5|^2 + |g^\prime_6|^2
\right) \alpha_2 (x_1,x_2) \ ,
\end{eqnarray}
\begin{eqnarray}
 Y^{\prime \prime} &=&
\left(
\frac{|g^\prime_1|^2}{16} - \frac{|g^\prime_2|^2}{16} 
+|g^\prime_3|^2 -|g^\prime_4|^2
\right) \alpha_1 (x_1,x_2)
+\left(
|g^\prime_5|^2-|g^\prime_6|^2
\right) \beta_1 (x_1,x_2) \ ,
\end{eqnarray}
\begin{eqnarray}
 Z^{\prime \prime} &=&
\left(
|g^\prime_5|^2 - |g^\prime_6|^2
\right) \gamma_1 (x_1,x_2)
\ ,
\end{eqnarray}
where $\alpha_{1-2}$, $\beta_1$, and $\gamma_1$ are
the same functions that we defined in $\tau^+ \to \mu^+ \mu^+ \mu^-$
calculation.
$X^{\prime \prime}$,
$Y^{\prime \prime}$, and $Z^{\prime \prime}$
can be extracted by
asymmetric integrations as before,
but we cannot obtain information on CP violation in this case.

Notice that the above three cases exhaust all possibilities 
in the three body decay of $\tau$ to $e$ and/or $\mu$
as long as we neglect the electron and muon masses compared to 
the $\tau$ mass.
Namely, the formula for other cases can be
obtained by appropriate replacements of $e$ and/or $\mu$.

The formulae for LFV decays with $\tau^-$ can be
obtained in a similar substitution as the $\tau \to \mu \gamma$ case.
Using appropriate angles of $\tau^-$ decay in Frame 3
and $\tau^+$ decay in Frame 2,
$dR_b$ gets an extra minus sign in eqs.(\ref{54}), (\ref{neo66}),
and (\ref{73}).

\section{$\tau \to \mu \nu \bar{\nu} \gamma$ process and
background suppressions}

In this section,
we consider the 
background processes
for the $\tau \to \mu \gamma$ search,
and we show that the measurement of
angular distributions is useful in 
identifying the background process.
In the muon decay,
the physical background can be suppressed 
if we use polarized muons \cite{Kuno:1996kv}.
In the following,
we show a similar suppression mechanism
holds for $\tau$ decay
if we use the spin correlation.

One of the main background for the
$\tau \to \mu \gamma$ search comes from
the kinematical endpoint region of 
the $\tau \to \mu \nu \bar{\nu} \gamma$ process
where two neutrinos carry out a little energy
at the rest frame of $\tau$.
In the following,
we assume that $\tau^+$ decays into 
$\mu^+ \nu \bar{\nu} \gamma$ and 
$\tau^-$ decays through one of
hadronic and leptonic decay processes.
For the $\tau^-$ decay,
the differential branching ratio
and the spin dependence term are 
given in eqs.(\ref{28})--(\ref{35}).
For $\tau^+ \to \mu^+ \nu \bar{\nu} \gamma$,
these quantities are given by
\begin{eqnarray}
 dB^{\rm B.G.} =
\frac{1}{\Gamma}\ 
\frac{G_{\rm F}^2 m_\tau^5 \alpha}{3 \times 2^{11} \pi^5}\ 
dx\ dy\ dz\ d \Omega_\mu \  \sin z\ \frac{\beta_\mu}{y}\ F
\ ,
\end{eqnarray}
\begin{eqnarray}
 dR^{\rm B.G.}_b \!\!\!&=&\!\!\!
\frac{1}{\Gamma}\ 
\frac{G_{\rm F}^2 m_\tau^5 \alpha}{3 \times 2^{11} \pi^5}\ 
dx\ dy\ dz\ d \Omega_\mu \  \sin z\ \frac{\beta_\mu}{y}\ 
\nonumber \\
&&\ \ \ \ \  \times
\left( -\beta_\mu G + H \cos z\ \right)
\left(
\begin{array}{c}
 \sin \theta_\mu \cos \phi_\mu \\
 \sin \theta_\mu \sin \phi_\mu \\
 \cos \theta_\mu \\
\end{array}
\right)
\ ,
\end{eqnarray}
where $x$ and $y$ are the muon and photon energies
normalized by $m_\tau/2$, respectively,
and 
($\theta_\mu$, $\phi_\mu$) is the polar coordinate
of the unit vector of the muon momentum direction,
all defined in the rest frame of $\tau^+$ (Frame 2).
$\beta_\mu = \sqrt{1-4r/x^2}$ with $r \equiv m_\mu^2/m_\tau^2$.
The angle $z$ is defined by 
$z \equiv \pi - \theta_{\mu \gamma}$,
where $\theta_{\mu \gamma}$ is the angle between
the muon and photon momentum in the same frame.
These quantities can be obtained by a simple 
replacement from the formula of the 
differential decay width for the radiative muon decay
presented in Ref.\cite{Kuno:1999jp}.
For completeness,
the functions $F$, $G$, and $H$ are given in Appendix \ref{functions}.

The background comes from 
the kinematical region near
$x=1+r$ and $y=1-r$,
at which the branching fraction vanishes.
However,
with finite detector resolutions,
this kinematical region gives physical backgrounds.
If we take the signal region as
$1+r-\delta x \leq x \leq 1+r$ and $1-r-\delta y \leq y \leq 1-r$,
the leading terms of 
the branching ratio and spin dependence term
expanded in terms of 
$r$, $\delta x$, and $\delta y$,
after integrating over $z$,
are given by
\begin{eqnarray}
 dB^{\rm B.G.} \simeq
\frac{1}{\Gamma}\ 
\frac{G_{\rm F}^2 m_\tau^5 \alpha}{3 \times 2^{11} \pi^5}\ 
d \Omega_\mu
\left(
\delta x^4 \delta y^2 + \frac{8}{3}\ \delta x^3 \delta y^3
\right)\ ,
\label{96}
\end{eqnarray}
\begin{eqnarray}
  dR^{\rm B.G.}_b \simeq
\frac{1}{\Gamma}\ 
\frac{G_{\rm F}^2 m_\tau^5 \alpha}{3 \times 2^{11} \pi^5}\ 
d \Omega_\mu
\left(
- \delta x^4 \delta y^2 + \frac{8}{3}\ \delta x^3 \delta y^3
\right) 
\left(
\begin{array}{c}
 \sin \theta_\mu \cos \phi_\mu \\
 \sin \theta_\mu \sin \phi_\mu \\
 \cos \theta_\mu \\
\end{array}
\right)
\ .
\label{97}
\end{eqnarray}
Then after integrating over $\phi_\mu$, $\phi_\pi$, $\phi_\tau$,
and $\theta_\tau$,
the differential cross section for 
$e^+ e^- \to \tau^+ \tau^- \to \mu^+ \nu \bar{\nu} \gamma + \pi^- \nu$
is given by
\begin{eqnarray}
 &&\!\!\!\!
d\sigma (e^+ e^- \to \tau^+ \tau^- \to \mu^+ \nu \bar{\nu} \gamma + \pi^- \nu)
\nonumber \\
&& = \sigma (e^+ e^- \to \tau^+ \tau^-) 
B (\tau^- \to \pi^- \nu) 
\left(
\frac{G_{\rm F}^2 m_\tau^5 \alpha}{3 \times 2^{9} \pi^4}
/ \Gamma
\right)
\frac{d \cos \theta_\mu}{2}\
\frac{d \cos \theta_\pi}{2} \nonumber \\
&& \hspace*{1cm} \times
\left \{
\left(
\delta x^4 \delta y^2 + \frac{8}{3} \delta x^3 \delta y^3
\right)
 - \frac{s-2m_\tau^2}{s+2 m_\tau^2}\ 
\left(
- \delta x^4 \delta y^2 + \frac{8}{3} \delta x^3 \delta y^3
\right)
\cos \theta_\mu \cos \theta_\pi
\right \} \ .
\nonumber \\
\end{eqnarray}
If the photon energy resolution is worse
than the muon energy resolution, 
the term $\delta x^4 \delta y^2$ is small
compared to $(8/3) \delta x^3 \delta y^3$.
In such a case,
the angular distribution is similar to the
$A_R=0$, $A_L \neq 0$ case of
the $\tau \to \mu \gamma$ angular distribution.
See eqs.(\ref{18}), (\ref{19}), and (\ref{41}).
This feature is useful for the 
background suppressions for $\tau^+ \to \mu^+_R \gamma$ search
because signal and background processes have
different angular correlation.
For $\tau^+ \to \mu^+_L \gamma$ search,
the signal to background ratio is almost the same
even if we take into account angular correlation.

A similar background suppression works for 
$\tau \to e \gamma$ case because eqs.(\ref{96}) and (\ref{97})
do not include the mass of the muon explicitly.

\section{Summary}

In this paper,
we have calculated
the differential cross sections
of $e^+ e^- \to \tau^+ \tau^- \to f_B f_A$ processes,
where one of $\tau$'s decays through LFV processes.
Using spin correlations of $\tau^+ \tau^-$,
we show that the P odd asymmetry of $\tau \to \mu \gamma$
and $\tau \to e \gamma$
and P and T asymmetries of three body LFV decays of $\tau$
can be obtained by angular correlations.
These P and T odd quantities are important to identify
a model of new physics responsible for LFV
processes.

We have also considered the
background suppression of the $\tau \to \mu \gamma$ and 
$\tau \to e \gamma$ search
by the angular distributions.
We see that
the analysis of the angular distributions
are useful for the $\tau^+ \to \mu^+_R \gamma$ 
($\tau^- \to \mu^-_L \gamma$) and 
$\tau^+ \to e^+_R \gamma$ 
($\tau^- \to e^-_L \gamma$) searches.

We can obtain similar information in muon decay experiments
if initial muons are polarized.
Although highly polarized muons are available experimentally,
a special setup for 
production and transportation of a muon beam
is necessary for actual experiment.
The advantage of the $\tau$ case is that
we can extract the information on $\tau$ spins by
looking at the decay distribution of the other side of 
$\tau$ decay so that
we do not need a special requirement for experimental setup.

\section*{Acknowledgments}

The authors would like to thank A. I. Sanda
for suggesting us to consider spin correlation
in the LFV decay of $\tau$.
They also would like to
thank J. Hisano for useful discussions and comments.

\appendix
\section{The derivation of the general formulae} \label{formula}

In this section,
we derive the eq.(\ref{neo1}) from
the amplitude in eq.(\ref{1}).

By using the completeness relation of the 
fermion spinors,
the amplitude squared is deformed to
\begin{eqnarray}
&& \!\!\!\!
\left| 
\bar{A} ( \psla_A + m_\tau ) \gamma^\mu ( \psla_B - m_\tau ) B
\right|^2 
\nonumber \\
\!\!\!&=&\!\!\!
\left|
\sum_{\lambda_1 = \pm} \sum_{\lambda_2 = \pm} 
\bar{A}  u ({\bf p}_A, \lambda_1)
\bar{u} ({\bf p}_A, \lambda_1) \gamma^\mu
v ({\bf p}_B, \lambda_2)
\bar{v} ({\bf p}_B, \lambda_2)
B
\right|^2
\nonumber \\
\!\!\!&=&\!\!\!
\sum_{\lambda_1 = \pm} \sum_{\lambda_2 = \pm} 
\sum_{\lambda_1^\prime = \pm} \sum_{\lambda_2^\prime = \pm} 
\left(
\bar{A} u ({\bf p}_A, \lambda_1) \bar{u} ({\bf p}_A, \lambda_1^\prime) A
\right)
\nonumber \\
&& \times
\left(
\bar{u} ({\bf p}_A, \lambda_1) \gamma^\mu
v ({\bf p}_B, \lambda_2)
\bar{v} ({\bf p}_B, \lambda_2^\prime) \gamma^\nu
u ({\bf p}_A, \lambda_1^\prime)
\right)
\left(
\bar{B} v ({\bf p}_B, \lambda_2^\prime) 
\bar{v} ({\bf p}_B, \lambda_2) B
\right)\ ,
\label{60}
\end{eqnarray}
where $\lambda$'s are the spin eigenvalues.
The spin summation can be performed by using
the Bouchiat-Michel formulae as follows \cite{bouchiat:1958,Haber:1994pe}:
\begin{eqnarray}
&&\!\!\!\!\!\!
\left| 
\bar{A} ( \psla_A + m_\tau ) \gamma^\mu ( \psla_B - m_\tau ) B
\right|^2 
\nonumber \\
\!\!\!&=&\!\!\!
\alpha^{D_-} \ 
{\rm Tr}\left[
(\psla_A + m_\tau) \gamma^\mu (\psla_B - m_\tau) \gamma^\nu
\right]\ 
\alpha^{D_+}
\nonumber \\
&&+\ 
\alpha^{D_-}\ 
{\rm Tr}\left[
(\psla_A + m_\tau ) \gamma^\mu \gamma_5 \ssla_B^b 
(\psla_B - m_\tau ) \gamma^\nu
\right]\ 
\rho_b^{D_+}
\nonumber \\
&&+\ 
\rho_a^{D_-}\ 
{\rm Tr}\left[
\gamma_5 \ssla_A^a
(\psla_A + m_\tau ) \gamma^\mu 
(\psla_B - m_\tau ) \gamma^\nu
\right]\ 
\alpha^{D_+}
\nonumber \\
&&+\ 
\rho_a^{D_-}\ 
{\rm Tr}\left[
\gamma_5 \ssla_A^a
(\psla_A + m_\tau ) \gamma^\mu \gamma_5 \ssla_B^b 
(\psla_B - m_\tau ) \gamma^\nu
\right]\ 
\rho_b^{D_+}
\ ,
\label{66}
\end{eqnarray}
where 
\begin{eqnarray}
 \alpha^{D_-} =
\frac{1}{2}\ 
\left \{
\bar{A} ( \psla_A + m_\tau ) A
\right \}\ ,
\ \ \ 
 \alpha^{D_+} =
\frac{1}{2}\ 
\left \{
\bar{B} ( \psla_B - m_\tau ) B
\right \}\ ,
\end{eqnarray}
\begin{eqnarray}
 \rho^{D_-}_a =
\frac{1}{2}\ 
\left \{
\bar{A} \ \gamma_5\  \ssla_A^a \ ( \psla_A + m_\tau ) A
\right \}\ ,
\ \ \ 
 \rho^{D_+}_b =
\frac{1}{2}\ 
\left \{
\bar{B} \ \gamma_5 \  \ssla_B^b \ ( \psla_B - m_\tau ) B
\right \}\ ,
\end{eqnarray}
where $(s_A^a)^\mu$ and $(s_B^b)^\nu$ are
four vectors which satisfy the following equations:
\begin{eqnarray}
&&  p_A \cdot s_A^a = p_B \cdot s_B^b = 0, \\
&&  s_A^a \cdot s_A^b = s_B^a \cdot s_B^b = - \delta^{ab} \ , 
\\
&&  \sum_{a=1}^3 (s_A^a)_\mu (s_A^a)_\nu = 
- g_{\mu \nu} 
+ \frac{p_{A \mu} p_{A \nu}}{m_\tau^2}\ ,\ \ \ 
\sum_{b=1}^3 (s_B^b)_\mu (s_B^b)_\nu = 
- g_{\mu \nu} 
+ \frac{p_{B \mu} p_{B \nu}}{m_\tau^2}\ .
\end{eqnarray}
The second and third terms
in eq.(\ref{66})
vanish because
the production parts are
antisymmetric on $\mu$ and $\nu$ indices
while the square of the electromagnetic current from
$e^+ e^-$ collision is symmetric on
$\mu$ and $\nu$ indices.
Explicit calculation gives
\begin{eqnarray}
 {\rm Tr}\left[
(\psla_A + m_\tau ) \gamma^\mu \gamma_5 \ssla_B^b 
(\psla_B - m_\tau ) \gamma^\nu
\right] =
4i m_\tau \epsilon^{\mu \nu \rho \sigma}
p_{B \rho} (s^b_{B})_\sigma
+
4i m_\tau \epsilon^{\mu \nu \rho \sigma}
p_{A \rho} (s^b_{B})_\sigma
\ ,
\end{eqnarray}
\begin{eqnarray}
 {\rm Tr}\left[
\gamma_5 \ssla_A^a 
(\psla_A + m_\tau ) \gamma^\mu 
(\psla_B - m_\tau ) \gamma^\nu
\right] =
4i m_\tau \epsilon^{\mu \nu \rho \sigma}
p_{B \rho} (s^a_{A})_\sigma
+
4i m_\tau \epsilon^{\mu \nu \rho \sigma}
p_{A \rho} (s^a_{A})_\sigma
\ .
\end{eqnarray}
\begin{eqnarray}
\sum_{\rm spin} | \bar{v}_{e^+} \gamma_\mu u_{e^-} |^2
\!\!\!&=&\!\!\!
{\rm Tr} \left[
\psla_{e^+} \gamma_\mu \psla_{e^-} \gamma_\nu
\right]\nonumber \\
\!\!\!&=&\!\!\!
4 p_{e^+ \mu} p_{e^- \nu} +
4 p_{e^+ \nu} p_{e^- \mu} -
4 g_{\mu \nu}
p_{e^+} \cdot p_{e^-} \ .
\end{eqnarray}
Using the narrow width approximation,
\begin{eqnarray}
 \left|
\frac{1}{q^2-\left(m-\frac{i\Gamma}{2} \right)^2}
\right|^2
\simeq \frac{\pi}{m \Gamma} \delta (q^2-m^2) \ ,
\label{58}
\end{eqnarray}
the first and last terms in eq.(\ref{66})
give formula (\ref{neo1}) 
after the phase space integral.

\section{The kinematical functions} \label{functions}

In this section,
we list the kinematical functions 
used in the formulae of branching ratios.

The functions $\alpha_{1-5}$, $\beta_{1-2}$, and $\gamma_{1-4}$
in the $\tau^+ \to \mu^+ \mu^+ \mu^-$ 
and $\tau^+ \to \mu^- e^+ e^+$
decay calculations are given as follows.
These functions are the same as those used 
in $\mu^+ \to e^+ e^+ e^-$ decay \cite{Okada:2000zk}.
$x_1$ and $x_2$ are given by $x_1 = 2E_1/m_\tau$
and $x_2 = 2 E_2 / m_\tau$.
\begin{eqnarray}
 \alpha_1 (x_1, x_2) = 8(2-x_1-x_2)(x_1+x_2-1)\ ,
\end{eqnarray}
\begin{eqnarray}
 \alpha_2 (x_1,x_2) = 2 \left \{ 
x_1 (1- x_1) + x_2 (1- x_2)
\right \}\ ,
\end{eqnarray}
\begin{eqnarray}
 \alpha_3 (x_1,x_2)= 8 \left \{ 
\frac{2 x_2^2 -2 x_2 +1 }{1-x_1}
+ \frac{2 x_1^2 - 2 x_1 + 1}{1-x_2}
\right \}\ ,
\end{eqnarray}
\begin{eqnarray}
 \alpha_4 (x_1,x_2)= 32 (x_1 + x_2 -1 )\ ,
\end{eqnarray}
\begin{eqnarray}
 \alpha_5 (x_1,x_2) = 8 (2 -x_1 -x_2)\ ,
\end{eqnarray}
\begin{eqnarray}
 \beta_1 (x_1,x_2) = 
\frac{ 2 (x_1 +x_2)(x_1^2 +x_2^2)-6 (x_1+x_2)^2 +12(x_1 + x_2) -8}
{2 - x_1 - x_2}
\end{eqnarray}
\begin{eqnarray}
 \beta_2 (x_1,x_2) \!\!\!&=&\!\!\!
\frac{ 8}{(1-x_1)(1-x_2)(2 -x_1-x_2)}
\nonumber \\
&& \times \left \{ {\rule[-3mm]{0mm}{10mm}\ }
2 (x_1+x_2)(x_1^3+x_2^3) - 4 (x_1+x_2)(2 x_1^2 +x_1 x_2 + 2 x_2^2)
\right.
\nonumber \\
&& \left.
\ \ \ \ \ \ + (19x_1^2 +30x_1x_2 +19 x_2^2)
-12 (2 x_1+2x_2 -1) 
{\rule[-3mm]{0mm}{10mm}\ }
\right \}\ ,
\end{eqnarray}
\begin{eqnarray}
 \gamma_1 (x_1,x_2) =
\frac{ 4 \sqrt{(1-x_1)(1-x_2)(x_1+x_2-1)} (x_2-x_1)}{2 - x_1 -x_2}\ ,
\end{eqnarray}
\begin{eqnarray}
 \gamma_2 (x_1,x_2) =
32 \sqrt{ \frac{ x_1+x_2-1}{(1-x_1)(1-x_2)} }\ 
\frac{(x_1+x_2-1)(x_2-x_1)}{2-x_1-x_2}\ ,
\end{eqnarray}
\begin{eqnarray}
 \gamma_3 (x_1,x_2) =
16 \sqrt{ \frac{ x_1+x_2-1}{(1-x_1)(1-x_2)} }\ 
(x_1 + x_2 -1 )(x_2 -x_1)\ ,
\end{eqnarray}
\begin{eqnarray}
 \gamma_4 (x_1,x_2) =
8 \sqrt{ \frac{ x_1+x_2-1}{(1-x_1)(1-x_2)} }\ 
(2 - x_1 - x_2 )(x_2-x_1)\ .
\end{eqnarray}

The functions $A_{1-7}$, $B_{1-4}$, and $C_{1-6}$
in the $\tau^+ \to \mu^+ e^+ e^-$ decay calculation are given by
\begin{eqnarray}
A_1(x_1,x_2)= \frac{8\,\left( 2 - x_1 - 4\,x_2 + 2\,x_1\,x_2 + 
      2\,x_2^2 \right) }{1 - x_1} \ ,
\end{eqnarray}
\begin{eqnarray}
 A_2(x_1,x_2)=-8\,\left(  x_1 + x_2 -1 \right) \ ,
\end{eqnarray}
\begin{eqnarray}
 A_3(x_2)=-8\,\left( 1 - x_2 \right)\ ,
\end{eqnarray}
\begin{eqnarray}
 A_4(x_1)=\frac{x_1 \left( 1 - x_1 \right)  }{2}\ ,
\end{eqnarray}
\begin{eqnarray}
 A_5(x_1,x_2)=2\,\left( 2 - x_1 - x_2 \right) \,
  \left( x_1 + x_2 -1 \right)\ ,
\end{eqnarray}
\begin{eqnarray}
 A_6(x_2)=2\,x_2 \left( 1 - x_2 \right) \ ,
\end{eqnarray}
\begin{eqnarray}
 A_7(x_1,x_2)=-8\,\left( 4 - 5\,x_1 + x_1^2 - 8\,x_2 + 
    4\,x_1\,x_2 + 4\,x_2^2 \right)\ ,
\end{eqnarray}
\begin{eqnarray}
 A_8(x_1,x_2)=-4\,\left( 1 - x_1 \right) \,
  \left( x_1 + 2\,x_2 -2\right)\ ,
\end{eqnarray}
\begin{eqnarray}
 B_1(x_1,x_2) \!\!\!&=&\!\!\!
\frac{-8}{\left( 1 - x_1 \right) \,
    \left( 2- x_1 - x_2\right) } 
\nonumber \\
&& \times
\left( -6 + 8x_1 - 3x_1^2 + 12x_2 - 
      11 x_1 x_2 + 2 x_1^2 x_2 - 
      8 x_2^2 + 4 x_1 x_2^2 + 2 x_2^3
      \right)\ ,
\nonumber \\
\end{eqnarray}
\begin{eqnarray}
 B_2(x_1,x_2)=\frac{- \left( 1 - x_1 \right) \,
      \left( 2 - 2\,x_1 + x_1^2 - 2\,x_2 + 
        x_1\,x_2 \right)  }{2\,
    \left( 2 - x_1 - x_2 \right) }\ ,
\end{eqnarray}
\begin{eqnarray}
 B_3(x_1,x_2)=\frac{2\,\left( 1 - x_2 \right) \,
    \left( 2 - 2\,x_1 - 2\,x_2 + x_1\,x_2 + 
      x_2^2 \right) }{2-x_1 - x_2}\ ,
\end{eqnarray}
\begin{eqnarray}
 B_4(x_1,x_2) \!\!\!&=&\!\!\! 
\frac{8}{2 - x_1 - x_2}
\nonumber \\
&& \times
\left( {\rule[-3mm]{0mm}{5mm}\ } \right.
-10 + 16x_1 - 7x_1^2 + x_1^3 + 
      22x_2 - 23x_1x_2
\nonumber \\
&&
\hspace*{2cm} + 5x_1^2x_2 - 16x_2^2 + 
      8x_1x_2^2 + 4x_2^3 
\left. {\rule[-3mm]{0mm}{5mm}\ } \right)\  ,
\end{eqnarray}
\begin{eqnarray}
 B_5(x_1,x_2)=\frac{4\,\left( 1 - {x_1} \right) \,
    \left( 2 - 4\,{x_1} + {{x_1}}^2 - 4\,{x_2} + 
      3\,{x_1}\,{x_2} + 2\,{{x_2}}^2 \right) }
     {2 - {x_1} - {x_2}}\ ,
\end{eqnarray}
\begin{eqnarray}
 C_1(x_1,x_2)=\frac{-16\,\left( x_1 + x_2 -1 \right) \,
    {\sqrt{\left( 1 - x_1 \right) \,\left( 1 - x_2 \right) \,
        \left( x_1 + x_2 -1 \right) }}}{\left( 1 - x_1
      \right) \,\left( 2 - x_1 - x_2 \right) }\ ,
\end{eqnarray}
\begin{eqnarray}
 C_2(x_1,x_2)=\frac{8\,\left( x_1 + x_2 -1 \right) \,
    {\sqrt{\left( 1 - x_1 \right) \,\left( 1 - x_2 \right) \,
        \left( x_1 + x_2 -1 \right) }}}{1 - x_1}\ ,
\end{eqnarray}
\begin{eqnarray}
 C_3(x_1,x_2)=\frac{-8\,\left( 1 - x_2 \right) \,
    {\sqrt{\left( 1 - x_1 \right) \,\left( 1 - x_2 \right) \,
        \left( x_1 + x_2 -1 \right) }}}{1 - x_1}\ ,
\end{eqnarray}
\begin{eqnarray}
 C_4(x_1,x_2)=\frac{\left( 1 - x_1 \right) \,
    {\sqrt{\left( 1 - x_1 \right) \,\left( 1 - x_2 \right) \,
        \left( x_1 + x_2 -1 \right) }}}{2 - x_1 - x_2}\ ,
\end{eqnarray}
\begin{eqnarray}
 C_5(x_1,x_2)=\frac{4\,\left( 1 - x_2 \right) \,
    {\sqrt{\left( 1 - x_1 \right) \,\left( 1 - x_2 \right) \,
        \left( x_1 + x_2 -1 \right) }}}{2 - x_1 - x_2}\ ,
\end{eqnarray}
\begin{eqnarray}
 C_6(x_1,x_2)=\frac{-16\,{\sqrt{\left( 1 - x_1 \right) \,
\left( 1 - x_2 \right) \,
        \left( x_1 + x_2 -1 \right) }}\,
    \left( 3 - x_1 - 2\,x_2 \right) }{2 - x_1 - x_2}\ ,
\end{eqnarray}
\begin{eqnarray}
 C_7(x_1,x_2)=8\,{\sqrt{\left( 1 - {x_1} \right) \,
      \left( 1 - {x_2} \right) \,
      \left( {x_1} + {x_2} -1 \right) }}\ .
\end{eqnarray}

Finally,
the functions $F$, $G$, and $H$ in
the $\tau \to \mu \nu \bar{\nu} \gamma$ decay calculation
are given by
\begin{eqnarray}
 F = F^{(0)} + r F^{(1)} + r^2 F^{(2)}\ ,
\end{eqnarray}
\begin{eqnarray}
 G = G^{(0)} + r G^{(1)} + r^2 G^{(2)}\ ,
\end{eqnarray}
\begin{eqnarray}
 H = H^{(0)} + r H^{(1)} + r^2 H^{(2)}\ ,
\end{eqnarray}
where 
$F^{(0)-(2)}$, $G^{(0)-(2)}$, and $H^{(0)-(2)}$
are the functions of $x ( \equiv 2 E_\mu / m_\tau )$, 
$y ( \equiv 2 E_\gamma / m_\tau)$,  
$d (\equiv 1 + \beta_\mu \cos z)$
with $\beta_\mu = \sqrt{ 1 - 4r/x^2}$
($r \equiv m_\mu^2/m_\tau^2$)
and $z = \pi - \theta_{\mu \gamma}$.
These functions are given by
\begin{eqnarray}
 F^{(0)}(x,y,d) \!\!\!&=&\!\!\!
\frac{-8\,\left( -3 + 2\,x + 2\,y \right) \,
     \left( 2\,x^2 + 2\,x\,y + y^2 \right) }{d} 
\nonumber \\
&& + 
  8\,x\,\left \{ x^2\,\left( 2 + 4\,y \right)  + 
     y\,\left( -3 + y + y^2 \right)  + 
     x\,\left( -3 + y + 4\,y^2 \right) \right \}
\nonumber \\
&& - 
  2\,x^2\,y\,\left \{ -6 + y\,\left( 5 + 2\,y \right)  + 
     2\,x\,\left( 4 + 3\,y \right)  \right \} \,d 
\nonumber \\
&& + 
  2\,x^3\,y^2\,\left( 2 + y \right) \,d^2 
\ ,
\end{eqnarray}
\begin{eqnarray}
 F^{(1)}(x,y,d) \!\!\!&=&\!\!\!
\frac{32\,\left( x + y \right) \,
     \left( -3 + 2\,x + 2\,y \right) }{x\,d^2} + 
  \frac{8\,\left \{ 6\,x^2 + \left( 6 - 5\,y \right) \,y - 
       2\,x\,\left( 4 + y \right)  \right \} }{d} 
\nonumber \\
&& - 
  8\,x\,\left \{ -4 - \left( -3 + y \right) \,y + 
     3\,x\,\left( 1 + y \right)  \right \}  + 
  6\,x^2\,y\,\left( 2 + y \right) \,d
\ ,
\end{eqnarray}
\begin{eqnarray}
 F^{(2)}(x,y,d) \!\!\!&=&\!\!\!
\frac{-32\,\left( -4 + 3\,x + 3\,y \right) }{x\,d^2} + 
  \frac{48\,y}{d}
\ ,
\end{eqnarray}
\begin{eqnarray}
 G^{(0)} (x,y,d) \!\!\!&=&\!\!\!
\frac{-8\,x\,\left \{ 4\,x^2 + 
       y\,\left( -1 + 2\,y \right)  + 
       x\,\left( -2 + 6\,y \right)  \right \} }{d} 
\nonumber \\
&& + 
  4\,x^2\,\left \{ -2 + 3\,y + 4\,y^2 + 
     x\,\left( 4 + 6\,y \right)  \right \}  - 
  4\,x^3\,y\,\left( 2 + y \right) \,d 
\ ,
\end{eqnarray}
\begin{eqnarray}
G^{(1)} (x,y,d) \!\!\!&=&\!\!\!
\frac{32\,\left( -1 + 2\,x + 2\,y \right) }{d^2} + 
  \frac{8\,x\,\left( 6\,x - y \right) }{d} - 
  12\,x^2\,\left( 2 + y \right) 
\ ,
\end{eqnarray}
\begin{eqnarray}
 G^{(2)} (x,y,d) \!\!\!&=&\!\!\!
\frac{-96}{d^2}
\ ,
\end{eqnarray}
\begin{eqnarray}
 H^{(0)} (x,y,d) \!\!\!&=&\!\!\!
\frac{-8\,y\,\left( x + y \right) \,
     \left( -1 + 2\,x + 2\,y \right) }{d} 
\nonumber \\
&&
+ 
  4\,x\,y\,\left \{ 2\,x^2 + 2\,y\,\left( 1 + y \right)  + 
     x\,\left( -1 + 4\,y \right)  \right \}  
\nonumber \\
&& - 
  2\,x^2\,y^2\,\left( -1 + 4\,x + 2\,y \right) \,d + 
  2\,x^3\,y^3\,d^2 
\ ,
\end{eqnarray}
\begin{eqnarray}
 H^{(1)} (x,y,d) \!\!\!&=&\!\!\!
\frac{32\,y\,\left( -1 + 2\,x + 2\,y \right) }{x\,d^2} - 
  \frac{8\,y\,\left( -2 + x + 5\,y \right) }{d} 
\nonumber \\
&& - 
  4\,x\,\left( 3\,x - 2\,y \right) \,y + 6\,x^2\,y^2\,d
\ ,
\end{eqnarray}
\begin{eqnarray}
 H^{(2)} (x,y,d) \!\!\!&=&\!\!\!
\frac{-96\,y}{x\,d^2} + \frac{48\,y}{d}
\ .
\end{eqnarray}

\end{document}